\documentstyle[sprocl]{article}
\input{psfig}

\catcode`\@=11

\def\mycite{\@ifnextchar [{\@tempswatrue\@mycitex}{\@tempswafalse\@mycitex[]}}
\def\mcite{\@ifnextchar [{\@tempswatrue\@mycitex}{\@tempswafalse\@mycitex[]}}

\def\@mycitex[#1]#2{\if@filesw\immediate\write\@auxout{\string\citation{#2}}\fi
 \def\@citea{}\@mycite{\@for\@citeb:=#2\do
    {\@citea\def\@citea{,\penalty\@m\ }\@ifundefined
       {b@\@citeb}{{\bf ?}\@warning
       {Citation `\@citeb' on page \thepage \space undefined}}%
\hbox{\csname b@\@citeb\endcsname}}}{#1}}

\def\@mycite#1{[{#1}]}

\catcode`\@=12
\def\zbc{3.1.1}
\def\pbt{3.1.2}
\def\oat{3.1.3}
\def\bps{5.1.1}
\def\bzt{5.1.2}
\def\bzf{5.1.3}
\def\bzs{5.1.4}
\def\ppb{5.1.5}
\def\bsdf{5.1.6}
\def\spp{5.2.1}
\def\szt{5.2.2}
\def\szf{5.2.3}
\def\szs{5.2.4}
\def\sczz{5.2.5}
\def\tds{7.1.2}
\def\lsm{7.1.1}
\def\mog{7.2.1}
\def\mom{7.2.2}
\def\molf{7.2.3}
\def\motf{7.2.4}
\def\rss{7.3.1}
\def\rgs{7.3.2}

\newcommand{\NP}{{\em Nucl.\ Phys.\ }}
\newcommand{\PL}{{\em Phys.\ Lett.\ }}
\newcommand{\PR}{{\em Phys.\ Rev.\ }}
\newcommand{\PRP}{{\em Phys.\ Rep.\ }}
\newcommand{\CMP}{{\em Comm.\ Math.\ Phys.\ }}
\newcommand{\MPL}{{\em Mod.\ Phys.\ Lett.\ }}
\newcommand{\PRL}{{\em Phys.\ Rev.\ Lett.\ }}
\newcommand{\IJMP}{{\em Int.\ J.\ Mod.\ Phys.\ }}

\def\gym{g_{\rm YM}}
\def\identity{{\rlap{1} \hskip 1.6pt \hbox{1}}}
\def\str{{\rm STr} \,}
\def\itt{{\cal  T}}
\def\ijj{{\cal J}}
\def\imm{{\cal M}}

\font\blackboard=msbm10 at 11pt
\font\blackboards=msbm7
\font\blackboardss=msbm5
\newfam\black
\textfont\black=\blackboard
\scriptfont\black=\blackboards
\scriptscriptfont\black=\blackboardss
\def\bb#1{{\fam\black\relax#1}}

\newcommand{\bz}{{\bb Z}}
\newcommand{\br}{{\bb R}}
\newcommand{\bc}{{\bb C}}

\newcommand{\junk}[1]{}

\def\thefootnote{\fnsymbol{footnote}}

\begin{document}
\begin{flushright}
PUPT-1762\\
hep-th/9801182
\end{flushright}
\vspace{0.4in}

\title{Lectures on D-branes, Gauge Theory and
M(atrices)\footnote{
Based on lectures given at the Trieste summer school on particle
physics and cosmology, June 1997}}
\author{Washington Taylor IV}
\address{Department of Physics,\\
Princeton University,\\
Princeton, New Jersey 08544,\\
U.S.A.\\
\tt wati@Princeton.edu}

%
%

\vspace{0.2in}

\maketitle \abstracts{These notes give a pedagogical introduction to
D-branes and Matrix theory.  The development of the material is
based on super Yang-Mills theory, which is the low-energy field theory
describing multiple D-branes.  The main goal of these notes is to
describe physical properties of D-branes in the language of Yang-Mills
theory, without recourse to string theory methods.  This approach is
motivated by the philosophy of Matrix theory, which asserts that all
the physics of light-front M-theory can be described by an appropriate super
Yang-Mills theory.  }

\vspace{0.6in}

\noindent
Table of Contents\\
{\bf
1. Introduction\\
2. D-branes and Super Yang-Mills Theory\\
3. D-branes and Duality ({\rm T-duality and S-duality in D-brane SYM theory})\\
4. Branes and Bundles ({\rm Constructing $(p \pm 2k)$-branes from
$p$-branes})\\ 
5. D-brane Interactions\\
6. M(atrix) theory: The Conjecture\\
7. Matrix theory: Symmetries, Objects and Interactions\\
8. Matrix theory: Further Developments\\
9. Conclusions}
\vspace{0.3in}

\def\thefootnote{\alph{footnote}}

\newpage
\section{Introduction}
\label{sec:introduction}

\subsection{Orientation}
{\junk{testing}} 
In the last several years there has been a revolution
in string theory.  There are two major developments responsible for
this revolution.
\vspace{0.05in}

\noindent
{\sl i}.  It has been found that
all five string theories, as well as 11-dimensional supergravity,
are related by duality symmetries and
seem to be aspects of one underlying theory whose fundamental
principles have not yet been elucidated.

\noindent
{\sl ii}.  String theories contain Dirichlet $p$-branes, also known as
``D-branes''.  These objects have been shown to play a fundamental
role in nonperturbative string theory.
\vspace{0.05in}

Dirichlet $p$-branes are dynamical objects which are extended in $p$ spatial
dimensions.  Their low-energy physics can be described by
supersymmetric gauge theory.  The goal of these lectures is to
describe the physical properties of D-branes
which can be understood from this Yang-Mills theory description.  There is a
two-fold motivation for taking this point of view.  At the superficial
level, super Yang-Mills theory describes much of the interesting
physics of D-branes, so it is a nice way of learning something about
these objects without having to know any sophisticated string theory
technology.  At a deeper level, there is a growing body of evidence
that super Yang-Mills theory contains far more information about
string theory than one might reasonably expect.  In fact, the recent
Matrix theory conjecture \mcite{BFSS} essentially states that the
simplest possible super Yang-Mills theory with 16 supersymmetries,
namely ${\cal N} = 16$ super Yang-Mills theory in 0 + 1 dimensions,
completely reproduces the physics of eleven-dimensional supergravity
in light-front gauge.

The point of view taken in these lectures is that many interesting
aspects of string theory can be derived from Yang-Mills theory.  This
is a theme which has been developed in a number of contexts in recent
research.  Conversely,
one of the other major themes of recent
developments in formal high-energy theory has been the idea that
string theory can tell us remarkable things about low-energy field
theories such as super Yang-Mills theory, particularly in a
nonperturbative context.  In these lectures we will not discuss any
results of the latter variety; however, it is useful to keep in mind the
two-way nature of the relationship between string theory and
Yang-Mills theory.

The body of knowledge related to D-branes and Yang-Mills theory is by
now quite enormous and is growing steadily.  Due to limitations on
time, space and the author's knowledge there are many interesting
developments which cannot be covered here.  As always in an effort of
this sort, the choice of topics covered largely reflects the prejudices of the
author.  An attempt has been made, however, to concentrate on a
somewhat systematic development of those concepts which are useful in
understanding recent progress in Matrix theory.  For a comprehensive
review of D-branes in string theory, the reader is referred to the
reviews of Polchinski et al. \mcite{pcj,Polchinski-TASI}.

These lectures begin with a review of how the low-energy Yang-Mills
description of D-branes arises in the context of string theory.  After
this introduction, we take super Yang-Mills theory as our starting
point and we proceed to discuss a number of aspects of D-brane and string
theory physics from this point of view.  In the last
lecture we  use the technology developed in the first four
lectures to discuss the recently developed Matrix theory.

\subsection{D-branes from string theory}

We now give a brief review of the manner in which D-branes appear in
string theory.  In particular, we give a heuristic description of how
supersymmetric Yang-Mills theory arises as a low-energy description of
parallel D-branes.  The discussion here is rather abbreviated; the
reader interested in further details is referred to the reviews of
Polchinski et al. \mcite{pcj,Polchinski-TASI} or to the original papers
mentioned below.
\begin{figure}
\vspace{-0.3in}
\psfig{figure=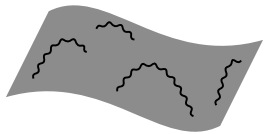,height=1.5in}
\vspace{-0.3in}
\caption[x]{\footnotesize D-branes (gray) are 
extended objects on which strings (black) can end}
\label{f:D-branes}
\end{figure}

In string theory,  Dirichlet $p$-branes are defined as $(p+1)$-dimensional
hypersurfaces in space-time on which strings are allowed to end (see
Figure~\ref{f:D-branes}).  From the point of view of perturbative
string theory, the positions of the D-branes are fixed, corresponding
to a particular string theory background.  The massless modes
of the open strings connected to the D-branes can be associated with
fluctuation modes of the D-branes themselves, however, so that in a full
nonperturbative context the D-branes are expected to become dynamical
$p$-dimensional membranes.  This picture is analogous to
the way in which, in a particular metric background for perturbative
string theory, the quantized closed string has massless graviton modes
which provide a mechanism for fluctuations in the metric itself.

The spectrum of low-energy fields in a given string background can be
simply computed from the string world-sheet field theory \mcite{GSW}.
Let us briefly review the analyses of the spectra for the string
theories in which we will be interested.  We consider two types of
strings: open strings, with endpoints which are free to move
independently, and closed strings, with no endpoints.  A superstring
theory is defined by a conformal field theory on the  $(1+1)$-dimensional
string world-sheet, with free bosonic fields $X^\mu$ corresponding to
the position of the string in 10 space-time coordinates, and fermionic
fields $\psi^\mu$ which are partners of the fields $X^\mu$ under
supersymmetry.  Just as for the classical string studied in beginning
physics courses, the degrees of freedom on the open string correspond
to standing wave modes of the fields; there are twice as many
modes on the closed string, corresponding to right-moving and
left-moving waves.  The open string boundary conditions on the bosonic fields
$X^\mu$ can be Neumann or Dirichlet for each field separately.  When
all boundary conditions are Neumann the string endpoints move freely
in space.  When $9-p$ of the fields have
Dirichlet boundary conditions, the string endpoints are constrained to
lie on a $p$-dimensional hypersurface which corresponds to a D-brane.
Different boundary conditions can also be chosen for the fermion
fields on the string.  On the open string, boundary conditions
corresponding to integer and half-integer modes are referred to as
Ramond (R) and Neveu-Schwarz (NS) respectively.  For the closed
string, we can separately choose periodic or antiperiodic boundary
conditions for the left- and right-moving fermions.  These give rise to
four distinct sectors for the closed string: NS-NS, R-R, NS-R and
R-NS.

Straightforward quantization of either the open or closed superstring
theory leads to several difficulties: the theory seems to contain a
tachyon with $M^2 < 0$, and the theory is not supersymmetric from the
point of view of ten-dimensional space-time.  It turns out that both of
these difficulties can be solved by projecting out half of the states
of the theory.  For the open string theory, there are two choices of
how this GSO projection operation can be realized.  These two
projections are equivalent, however, so that there is a unique spectrum for the
open superstring.  For the closed string, on the other hand, one can
either choose the same projection in the left and right sectors, or
opposite projections.  These two choices lead to the physically
distinct IIA and IIB closed superstring theories, respectively.

From the point of view of 10D space-time, the massless fields arising
from quantizing the string theory and incorporating the GSO projection
can be characterized by their transformation properties under ${\rm
spin}(8)$ (this is the covering group of the group $SO(8)$ which
leaves a lightlike momentum vector invariant).  We will now simply
quote these results from 
\mycite{GSW}.  For the
open string, in the NS sector there is a vector field $A_\mu$,
transforming under the $8_v$ representation of ${\rm spin} (8)$ and in
the R sector there is a fermion $\psi$ in the $8_s$ representation.
The massless fields for the IIA and IIB closed strings in the NS-NS
and R-R sectors are given in the following table:
\vspace{0.08in}
\begin{center}
\begin{tabular}{|  l | c | c |} \hline
& NS-NS& R-R \\\hline
IIA & $g_{\mu \nu}, \phi, B_{\mu \nu}$  & $A^{(1)}_\mu, A^{(3)}_{\mu
\nu \rho}$\\\hline
IIB & $g_{\mu \nu}, \phi, B_{\mu \nu}$  & $A^{(0)},
A^{(2)}_{\mu \nu}, A^{(4)}_{\mu
\nu \rho \sigma}$\\\hline
\end{tabular}
\end{center}
\vspace{0.08in}
The IIA and IIB strings have the same fields in the NS-NS sector,
corresponding to the space-time metric $g_{\mu \nu}$, dilaton $\phi$
and antisymmetric tensor field $B_{\mu \nu}$.  In addition, each
closed string theory has a set of R-R fields.  For the IIA theory
there are 1-form and 3-form fields.  For the IIB theory there is a
second scalar field (the axion), a second 2-form field, and a 4-form
field $A^{(4)}$ which is self-dual.  The NS-NS and R-R fields all
correspond to space-time bosonic fields.  In both the IIA and IIB theories
there are also fields in the NS-R and R-NS sectors corresponding to
space-time fermionic fields.

Until recently, the role of the R-R fields in string theory was rather
unclear.  In one of the most important papers in the recent
string revolution \mcite{Polchinski}, however, it was pointed out by Polchinski
that D-branes are charge carriers for these fields.  Generally, a
Dirichlet $p$-brane couples to the R-R $(p + 1)$-form field through a
term of the form
\begin{equation}
\mu_p \int_{\Sigma_{(p + 1)}}A^{(p + 1)}
\end{equation}
where the integral is taken over the $(p + 1)$-dimensional
world-volume of the $p$-brane.

In type IIA theory there are Dirichlet $p$-branes with $p= 0, 2, 4, 6,
8$ and in type IIB there can be Dirichlet $p$-branes with $p = -1, 1,
3, 5,7, 9$.  The D-branes with $p > 3$ couple to the duals of the R-R
fields, and are thus magnetically charged under the corresponding R-R
fields.  For example, a Dirichlet 6-brane, with a 7-dimensional
world-volume, couples to the 7-form whose 8-form field strength is the
dual of the 2-form field strength of $A^{(1)}$.  Thus, the Dirichlet
6-brane is magnetically charged under the R-R vector field in IIA
theory.  The story is slightly more complicated for the Dirichlet
8-brane and 9-brane \mcite{Polchinski-TASI}; however, 8-branes and
9-branes will not appear in these lectures in any significant way.

In addition to the Dirichlet $p$-branes which appear in type IIA and
IIB string theory, there are also solitonic NS-NS 5-branes which
appear in both theories, which are magnetically charged under the
NS-NS two-form field $B_{\mu \nu}$.  In the remainder of these notes
$p$-branes which are not explicitly stated to be Dirichlet or NS-NS
are understood to be Dirichlet $p$-branes; we will also sometimes
use the notation D$p$-brane to denote a $p$-brane of a particular dimension.

It is interesting to see how the dynamical degrees of freedom of a
D-brane arise from the massless string spectrum in a fixed D-brane
background \mcite{dlp}.  In the presence of a D-brane, the open string
vector field $A_\mu$ decomposes into components parallel to and
transverse to the D-brane world-volume.  
Because the endpoints of the strings are tied to the world-volume of
the brane, we can interpret these massless fields in terms of a
low-energy field theory on the D-brane world-volume.  The $p + 1$ parallel
components of $A_{\mu}$ turn into a $U(1)$ gauge field $A_{\alpha}$ on
the world-volume, while the remaining $9-p$ components appear as
scalar fields $X^a$.   The fields $X^a$ describe fluctuations of the
D-brane world-volume in transverse directions.  In general throughout
these notes we will use $\mu, \nu, \ldots$ to denote 10D indices,
$\alpha, \beta, \ldots$ to denote $(p + 1)$-D indices on a D-brane
world-volume, and $a, b, \ldots$ to denote $(9-d)$-D transverse indices.

One way to learn about the low-energy dynamics of a D-brane is to find
the equations of motion for the D-brane which must be satisfied for
the open string theory in the D-brane background to be conformally
invariant.  Such an analysis was carried out by Leigh \mcite{Leigh}.
He showed that in a purely bosonic theory, the equations of motion
for a D-brane are precisely those of the action
\begin{equation}
S = - T_p  \int d^{p + 1} \xi
\;e^{-\phi} \;\sqrt{-\det (G_{\alpha \beta} + B_{\alpha \beta} + 2 \pi \alpha'
F_{\alpha \beta})  }
\label{eq:DBI}
\end{equation}
where $G$, $B$  and $\phi$ are the pullbacks of the 10D metric,
antisymmetric tensor and dilaton to the D-brane world-volume, while $F$ is the
field strength of the world-volume $U(1)$ gauge field $A_{\alpha}$.
This action can be verified by a perturbative string
calculation \mcite{Polchinski-TASI}, which also gives a precise
expression for the brane tension
\begin{equation}
\tau_p =\frac{T_p}{g} =
 \frac{1}{g\sqrt{\alpha'}}  \frac{1}{ (2 \pi \sqrt{\alpha'})^{p}} 
\end{equation}
where $g = e^{\langle \phi \rangle}$ is the string coupling, equal to
the exponential 
of the dilaton expectation value, and $\alpha'$ is related to the
string tension through
\begin{equation}
\frac{1}{2 \pi \alpha'}  = T_{{\rm string}}.
\end{equation}
The inverse string coupling appears because the leading string diagram
which contributes to the action (\ref{eq:DBI}) is a disk diagram.

In the full supersymmetric string theory, the action (\ref{eq:DBI})
must be extended to a supersymmetric Born-Infeld type action.  In
addition, there are Chern-Simons type terms coupling the D-brane gauge
field to the R-R fields, of which the leading term is the $\int A^{(p
+ 1)}$ term discussed above; we will discuss these terms in more
detail later in these notes.

If we make a number of simplifying assumptions, the form of the action
(\ref{eq:DBI}) simplifies considerably.  First, let us assume that the
background ten-dimensional space-time is flat, so that $g_{\mu \nu} =
\eta_{\mu \nu}$ (we use a metric with signature $-++ \cdots ++$).
Further, let us assume that the D-brane is approximately flat and that
we can identify the world-volume coordinates on the D-brane with $p+1$
of the ten-dimensional coordinates (the static gauge assumption).
Then, the pullback of the metric to the D-brane world-volume becomes
\begin{equation}
G_{\alpha \beta} \approx \eta_{\alpha \beta} + \partial_\alpha X^a
\partial_\beta X^a+{\cal O} \left((\partial X)^4 \right)
\end{equation}
If we make the further assumptions that $B_{\mu \mu}$ vanishes, and
that $2 \pi \alpha' F_{\alpha \beta}$ and $\partial_\alpha X^a$ are
small and of the same order, then we see that the low-energy D-brane
world-volume action becomes
\begin{equation}
S =-\tau_pV_p  -\frac{1}{4g_{{\rm YM}}^2}
 \int d^{p + 1} \xi
\left(F_{\alpha \beta} F^{\alpha \beta} +\frac{2}{(2 \pi \alpha')^2} 
\partial_\alpha X^a \partial^\alpha X^a\right) +{\cal O} (F^4)
\label{eq:action-expansion}
\end{equation}
where $V_p$ is the $p$-brane world-volume and the coupling $\gym$ is given
by
\begin{equation}
\gym^2 = \frac{1}{4 \pi^2 \alpha'^2 \tau_p} 
= \frac{g}{\sqrt{\alpha'}}  (2 \pi \sqrt{\alpha'})^{p-2}
\label{eq:ym-coupling}
\end{equation}
The second term in (\ref{eq:action-expansion}) is essentially just the
action for a $U(1)$ gauge theory in $p + 1$ dimensions with $9-p$
scalar fields.  In fact, after including fermionic fields $\psi$
the low-energy action for a D-brane becomes precisely the
supersymmetric $U(1)$ Yang-Mills theory in $p + 1$ dimensions which
arises from dimensional reduction of the $U(1)$ Yang-Mills theory in
10 dimensions with ${\cal N} = 1$ supersymmetry.  The action of this
10D theory is 
\begin{equation}
S = \frac{1}{\gym^2}  \int d^{10}\xi \; \left(
 -\frac{1}{4} F_{\mu \nu}F^{\mu \nu}
+ \frac{i}{2}  \bar{\psi} \Gamma^\mu \partial_{\mu} \psi \right)
\label{eq:SYM1}
\end{equation}

In the next section we will discuss supersymmetric Yang-Mills theories
of this type in more detail.  To conclude this introductory
discussion let us consider briefly the situation where we have a number of
distinct D-branes.
\begin{figure}
\vspace{-0.3in}
\psfig{figure=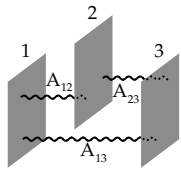,height=1.5in}
\vspace{-0.3in}
\caption[x]{\footnotesize  $U(N)$ fields $A_{ij}$ arise from strings stretched
between multiple D-branes}
\label{f:parallel-branes}
\end{figure}
In particular, let us imagine that we have $N$ parallel D-branes of
the same dimension, as depicted in Figure~\ref{f:parallel-branes}.  We
label the branes by an index $i$ running from $1$ to $N$.  There are
massless fields living on each D-brane world-volume, corresponding to
a gauge theory with total gauge group $U(1)^N$.  In addition, however,
we expect fields to arise corresponding to strings stretching between
each pair of branes.  These fields $A^{\mu}_{ij}$ carry 10D indices
$\mu$ as well as a pair of indices $i, j$ indicating which branes are
at the endpoints of the strings.  Because the strings are oriented,
there are $N^2 -N$ such fields (counting a vector $A_{\mu} =
(A_{\alpha}, X^a)$ as a single field).  The mass of a field
corresponding to a  string
connecting branes $i$ and $j$ is proportional to the distance between
these branes.  It was pointed out by Witten \mcite{Witten-bound} that
as the D-branes approach each other and the stretched strings become
massless, the fields arrange themselves precisely into the
gauge field components and adjoint scalars of a supersymmetric $U(N)$ gauge
theory in $p + 1$ dimensions.  Generally, such a super Yang-Mills
theory is described by the reduction to $p + 1$ dimensions of a 10D
non-abelian Yang-Mills theory where all fields are in
the adjoint representation of $U(N)$.

Thus, we see that with a number of simplifying assumptions, the
low-energy field theory describing a system of parallel D-branes is
simply a supersymmetric Yang-Mills (SYM) field theory.  In  the
following we will use SYM theory as the starting point from which to
analyze aspects of D-brane physics.

\section{D-branes and Super Yang-Mills Theory}

The previous section contained a fairly abbreviated discussion of the
string theory description of D-branes.  The most significant
part of this description for the purposes of these lectures is the
following statement, which we will treat as axiomatic in most of the sequel
\vspace{0.07in}

\noindent {\bf Starting point:}
The low-energy physics of $N$ Dirichlet $p$-branes living in
flat space is described in static gauge by the dimensional reduction to
$p + 1$ dimensions of
${\cal N} = 1$ SYM in 10D.
\vspace{0.05in}

In this section we fill in some of the details of this theory in  ten
dimensions, and describe explicitly the dimensionally
reduced theory in the case of 0-branes, $p = 0$.

\subsection{10D super Yang-Mills}

Ten-dimensional $U(N)$ super Yang-Mills theory has the action
\begin{equation}
S =\int d^{10}\xi \; \left( -\frac{1}{4} {\rm Tr}\; F_{\mu \nu}F^{\mu \nu}
+\frac{i}{2}  {\rm Tr}\; \bar{\psi} \Gamma^\mu D_{\mu} \psi \right)
\label{eq:SYM}
\end{equation}
where the field strength
\begin{equation}
F_{\mu \nu} = \partial_\mu A_\nu -\partial_\nu A_\mu -i g_{YM}[A_\mu, A_\nu]
\end{equation}
is the curvature of a $U(N)$ hermitian gauge field $A_\mu$.
The fields $A_\mu$ and $\psi$ are both in the adjoint representation
of $U(N)$ and carry adjoint indices which we will generally suppress.
The covariant derivative $D_\mu$ of $\psi$ is given by
\begin{equation}
D_\mu \psi = \partial_\mu \psi -i g_{\rm YM}[A_\mu, \psi]
\end{equation}
where $g_{{\rm YM}}$ is the Yang-Mills coupling constant.
$\psi$ is a 16-component Majorana-Weyl spinor of $SO(9,1)$.

The action (\ref{eq:SYM}) is invariant under the supersymmetry transformation
\begin{eqnarray}
\delta A_\mu & = & \frac{i}{2} \bar{\epsilon} \Gamma_\mu \psi \\
\delta \psi & = & -\frac{1}{4}  F_{\mu \nu} \Gamma^{\mu \nu} \epsilon
\label{eq:Yang-Mills-SUSY}
\nonumber
\end{eqnarray}
where $\epsilon$ is a  Majorana-Weyl spinor.  Thus, this
theory has 16 independent supercharges.  There are 8 on-shell bosonic
degrees of freedom and 8 fermionic degrees of freedom after imposition
of the Dirac equation.

Classically, this ten-dimensional super Yang-Mills action gives a well-defined
field theory.  The theory is anomalous, however, and therefore
problematic quantum mechanically.

It is often convenient to rescale the fields of the Yang-Mills
theory so that the coupling constant only appears as an overall
multiplicative factor in the action.  By absorbing a factor of $g_{\rm
YM}$ in $A$ and $\psi$, we find that the action is
\begin{equation}
S =\frac{1}{4g_{\rm YM}^2} \int d^{10}\xi \; \left( - {\rm Tr}\;
F_{\mu \nu}F^{\mu \nu} +2i {\rm Tr}\; \bar{\psi} \Gamma^\mu
D_{\mu} \psi \right)
\end{equation}
where the covariant derivative is given by
\begin{equation}
D_\mu = \partial_\mu -iA_\mu.
\end{equation}

\subsection{Dimensional reduction of super Yang-Mills}
%

The ten-dimensional  super Yang-Mills theory described in the previous
subsection can be used to construct a super Yang-Mills theory in $p +
1$ dimensions with 16 supercharges by the simple process of
dimensional reduction.  This is done by assuming that all fields are
independent of coordinates $p + 1, \ldots, 9$.  After dimensional
reduction, the  10D field $A_\mu$ decomposes into a $(p +
1)$-dimensional gauge field $A_\alpha$ and $9-p$ adjoint scalar fields
$X^a$.  The action of the dimensionally reduced theory takes the form
\begin{equation}
S = \frac{1}{4\gym^2}   \int d^{p + 1} \xi \;
{\rm Tr}\;(-F_{\alpha \beta} F^{\alpha \beta}-2(D_\alpha X^a)^2
+[X^a, X^b]^2
 +{\rm fermions}).
\label{eq:reduced}
\end{equation}

As discussed in Section \ref{sec:introduction}, this is precisely the
action describing the low-energy dynamics of $N$ coincident Dirichlet
$p$-branes in static gauge (although there the fields $X$ and $\psi$
are normalized
by the factor $X \rightarrow X/(2 \pi \alpha')$,
$\psi \rightarrow  \psi/(2 \pi \alpha')$).  The field
$A_\alpha$ is the gauge field on the D-brane world-volume, and the
fields $X^a$ describe transverse fluctuations of the D-branes.  Let us
comment briefly on the signs of the terms in the action
(\ref{eq:reduced}).  We would expect kinetic terms to appear with a
positive sign and potential terms to appear with a negative sign.
Because the metric we are using has a mostly positive signature, the
kinetic terms have a single raised 0 index corresponding to a change
of sign, so the kinetic terms indeed have the correct sign.  The
commutator term $[X^a, X^b]^2$ which acts as a potential term is
actually negative definite.  This follows from the fact that $[X^a,
X^b]^{\dagger} = [X^b, X^a]= -[X^a, X^b]$.  Thus, as expected, kinetic
terms in the action are positive while potential terms are
negative.

In order to understand the geometrical
significance of the fields $X^a$ it is useful to consider the field
configurations corresponding to classical vacua of the theory defined
by (\ref{eq:reduced}).  A classical vacuum corresponds to a static
solution of the equations of motion where the potential energy of the
system is minimized.  This occurs when the curvature $F_{\alpha \beta}$ and the
fermion fields vanish, and in addition the fields $X^a$ are covariantly
constant and commute with one another.  When the fields $X^a$ all
commute with one another at each point in  the $(p + 1)$-dimensional
world-volume of the branes, the fields can be simultaneously
diagonalized by a gauge transformation, so that we have
\begin{equation}
X^a = \left(\begin{array}{cccc}
x^a_1 & 0 &0 &  \ddots\\
0 & x^a_2 &\ddots & 0\\
0 & \ddots & \ddots & 0\\
 \ddots & 0 & 0 & x^a_N
\end{array}\right)
\end{equation}
In such a configuration, the $N$ diagonal elements of the matrix $X^a$
can be associated with the positions of the $N$ distinct D-branes in
the $a$-th transverse direction \mcite{Witten-bound}.  In accord with
this identification, one can easily verify that the masses of the
fields corresponding to off-diagonal matrix elements are precisely
given by the distances between the corresponding branes.

From this discussion, we see that the moduli space of classical vacua
for the $(p + 1)$-dimensional field theory arising from dimensional
reduction of 10D SYM is given by
\begin{equation}
\frac{(\br^{9-p})^N}{S_N} 
\end{equation}
The factors of $\br$ correspond to positions of the $N$ D-branes in
the $(9-p)$-dimensional transverse space.  The symmetry group $S_N$ is
the residual Weyl symmetry of the gauge group.  In the D-brane
language this corresponds to a permutation symmetry acting on the
D-branes, indicating that the D-branes should be treated as
indistinguishable objects.

As pointed out by Witten \mcite{Witten-bound}, a remarkable feature of
this description of D-branes is that an interpretation of a D-brane
configuration in terms of classical geometry can only be given when
the matrices $X^a$ are simultaneously diagonalizable.  In a generic
configuration, the positions of the D-branes are only roughly
describable through the spectrum of eigenvalues of the $X$ matrices.
This gives a natural and simple mechanism for the appearance of a
noncommutative geometry at short distances where the D-branes cease to
have well-defined positions according to classical commutative
geometry.

{}

\subsection{Example: 0-branes}
\label{sec:0-branes}

We now consider an explicit example of the dimensionally reduced
theory, that of the low-energy action of pointlike 0-branes.  This
system will be of central importance in the later sections on Matrix
theory.  As discussed above, the low-energy theory describing the
dynamics of $N$ 0-branes in a flat ten-dimensional space-time is the
dimensional reduction of ${\cal N} = 1$ 10D SYM to one space-time
dimension.  In the dimensionally reduced theory the ten-dimensional
vector field $A_\mu$ decomposes into 9 transverse scalars $X^a$, and a
1-dimensional gauge field $A_0$.  This theory has 16 supercharges, and
is therefore an ${\cal N} = 16$ supersymmetric matrix quantum
mechanics theory.  If we choose a gauge where the gauge field $A_0$
vanishes, then the Lagrangian for this theory is given by
\begin{eqnarray}
{\cal L} &=& \frac{1}{2 g \sqrt{\alpha'}} {\rm Tr}\; \left[
  \dot{X}^a \dot{X}_a
+\frac{1}{ (2 \pi \alpha')^2}
  \sum_{ a < b}[X^a, X^b]^2 \right. \label{eq:super-qm}\\
& &\hspace{1in}+  \left.
\frac{1}{2 \pi \alpha'} \theta^T i\dot{\theta}
- \frac{1}{ (2 \pi \alpha')^2}\theta^T \Gamma_a[X^a, \theta]) \right]\nonumber
\end{eqnarray}
Each of the  nine adjoint scalar matrices $X^a$ is a
hermitian $N \times N$ matrix, where $N$ is the number of 0-branes.
The superpartners of the $X$ fields are 16-component spinors $\theta$
which transform under the $SO(9)$ Clifford algebra given by the
$16 \times 16$ matrices $\Gamma^a$.  This theory was discussed many
years before the development of
D-branes \mcite{Claudson-Halpern,Flume,brr}; 
a more detailed discussion of this theory in the D-brane context can
be found in \mycite{dfs,Kabat-Pouliot,DKPS}.

%

The classical static solutions of this theory are found by minimizing
the potential, which occurs when $[X^a, X^b]= 0$ for all $a, b$.
As discussed in the general case, when the matrices can be
simultaneously diagonalized their 
diagonal elements can be interpreted geometrically as the coordinates
of the $N$ 0-branes.  The
classical configuration space of $N$ 0-branes is therefore given by
\begin{equation}
\frac{(\br^9)^N}{S_N}
\end{equation}
which is the configuration space of $N$ identical particles moving in
euclidean 9-dimensional space.
For a general configuration, the matrices cannot be diagonalized and the
off-diagonal elements only have a geometrical interpretation in terms
of a noncommutative geometry.

Note that for the 0-brane Yang-Mills theory, the reduction of the
original Born-Infeld theory is simpler than in higher dimensional
cases.  The only assumptions necessary to derive the 0-brane
Yang-Mills theory are that the background metric is flat and that the
velocities of the 0-branes are small.  Thus, the super Yang-Mills 0-brane
theory is essentially the nonrelativistic limit of the Born-Infeld
0-brane theory.  In the case of 0-branes, the assumption of static
gauge reduces to the assumption that there are no anti-0-branes in the
system.

\section{D-branes and Duality}
\label{sec:duality}

One of the most remarkable features of string theory is the intricate
network of duality symmetries relating the different consistent string
theories \mcite{Hull-Townsend,Witten-various}.  Such dualities
relate each of the five known superstring theories to one another and
to 11-dimensional supergravity

Some duality symmetries, such as the T-duality symmetry which relates
type IIA to type IIB, are perturbative symmetries; other dualities,
such as the S-duality symmetry of type IIB, are nonperturbative
symmetries which can take theories with a strong coupling to weakly
coupled theories.

Historically, D-branes were first studied using T-duality symmetry on
the string world-sheet \mcite{dlp}.  In this section we invert
the historical sequence of development and study duality symmetries
from the point of view of the low-energy field theories of D-branes.
We first discuss T-duality from the D-brane point of view.  We show
that without making reference to the string theory structure from
which it arose, the low-energy super Yang-Mills theory of $N$ D-branes
admits a T-duality symmetry when compactified on the torus.  We then
discuss S-duality of the IIB theory, which corresponds to super
Yang-Mills S-duality on the 3-brane world-volume.

\subsection{T-duality in super Yang-Mills theory}
\label{sec:T-duality}

Before deriving T-duality from the point of view of super Yang-Mills
theory, we briefly review what we expect of the type II T-duality
symmetry from string theory \mcite{Polchinski-TASI}.  T-duality is a
symmetry of type II string theory after one spatial dimension has been
compactified.  Let us compactify $X^9$ on a circle of radius $R_9$,
giving a space-time $\br^9 \times S^1$.  After such a
compactification, T-duality maps type IIA string theory compactified
on a circle of radius $R_9$ to type IIB string theory compactified on
a circle of radius $\hat{R}_9= \alpha'/R_9$.

On a string world-sheet, T-duality maps Neumann boundary conditions on
the bosonic field $X^9$ to Dirichlet boundary conditions and vice
versa.  Thus, for a fixed string background, T-duality maps
a $p$-brane to a $(p \pm 1)$-brane, where a brane originally wrapped
around the $X^9$ dimension is unwrapped by T-duality and vice versa.
This result in the context of perturbative string theory indicates
that we would expect the low-energy field theory of a system of
$p$-branes which are unwrapped in the transverse direction $X^9$ to be
equivalent to a field theory of $(p +1)$-branes wrapped on a dual circle
$\hat{X}^9$.  We now proceed to prove this result in a precise fashion,
using only the properties of the low-energy super Yang-Mills theory.
For the bulk of this subsection we set $2 \pi \alpha' = 1$ for
convenience;  constants are restored in the formulas at the end
of the discussion.  The arguments described in this subsection
originally appeared in
\mycite{WT-compact,BFSS,grt}.

\vspace{0.08in}
\noindent
{\it \zbc\ \ \ 0-branes on a circle}
\vspace{0.05in}

In order to simplify the discussion we begin with the simplest case,
corresponding to $N$ 0-branes moving on a space $\br^8 \times
S^1$.  The generalization to higher dimensional branes and to
T-dualities in multiple dimensions is straightforward and will be
discussed later.

As described in Section~\ref{sec:0-branes}, a system of $N$ 0-branes
moving in flat space $\br^9$  has a low-energy description in terms of
a supersymmetric matrix quantum mechanics.  The matrices in this
theory are $N \times N$ matrices, and the theory has 16 supersymmetry
generators.  In order to describe the motion of $N$ 0-branes in a
space where one direction is compactified, this theory must be
modified somewhat.  A naive approach would be to try to make the
matrices $X^9$ periodic.  This cannot be done without
increasing the number of degrees of freedom of the system, however.  One simple
way to see this is to note that the off-diagonal matrix elements
corresponding to strings stretching between different 0-branes have
masses proportional to the distance between the branes in the flat
space theory; this feature cannot be implemented in a compact
space without introducing an infinite number of degrees of freedom
corresponding to strings wrapping with an arbitrary homotopy class.

A systematic approach to describing the motion of 0-branes on $S^1$
can be developed along the lines of familiar orbifold techniques.  In
general, if we wish to describe the motion of $N$ 0-branes on a space
$\br^9/\Gamma$ which is the quotient of flat space by a discrete group
$\Gamma$, we can simply consider a system of $(N \cdot | \Gamma |)$
0-branes moving on $\br^9$ and then impose a set of constraints which
dictate that the brane configuration is invariant under the action of
$\Gamma$.  This approach was used by Douglas and Moore
\mcite{Douglas-Moore} to study the motion of 0-branes on spaces of the
form $\bc^2/\bz_k$; the authors showed that on such spaces the moduli
space of 0-brane configurations is modified quantum mechanically to
correspond to smooth ALE spaces.  Related work was done in
\mycite{Gimon-Polchinski,Johnson-Myers}.

\begin{figure}
\vspace{-0.3in}
\psfig{figure=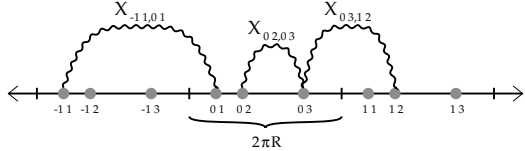,height=1.5in}
\vspace{-0.5in}
\caption[x]{\footnotesize  0-branes on the cover of  $S^1$ are indexed
by two integers}
\label{f:cover}
\end{figure}

In the case we are interested in here, the study of the motion of
0-branes in terms of a quotient space description is simplified since
there are no fixed points of the space under the action of any element
of the group $\Gamma$.  The universal covering space of $S^1$ is
$\br$, where $S^1 =\br/\bz$, so we can study 0-branes on $S^1$ by
considering the motion of an infinite family of 0-branes on $\br$.  If
we wish to describe $N$ 0-branes moving on $S^1$, then, we must
consider a family of 0-branes moving on $\br$ which are indexed by two
integers $n, i$ with $n \in{\bz}$ and $i \in\{1, \ldots, N\}$ (see
Figure~\ref{f:cover}).  This gives us a system described by $U
(\infty)$ matrix quantum mechanics with constraints.

The $U (\infty)$ theory describing the 0-branes on the covering space
has a set of matrix degrees of freedom described by fields $X^a_{mi,
nj}$.  Such a field corresponds to a string stretching from the $m$th
copy of 0-brane number $i$ to the $n$th copy of 0-brane number $j$.
For simplicity of notation, we will suppress the $i,j$ indices and
write these matrices as infinite matrices whose blocks $X^a_{mn}$ are
themselves $N \times N$ matrices.

The constraint of translation invariance under $\Gamma =\bz$ imposes
the condition that the theory is invariant under a simultaneous
translation of the $X^9$ coordinate by $2 \pi R_9$ and relabeling of the
indices $n$ by $n + 1$.  Mathematically, this condition says that
\begin{eqnarray}
X^a_{mn} & = & X^a_{(m-1)(n-1)},\;\;\;\;\;  a < 9 \nonumber\\
X^9_{mn} & = & X^9_{(m-1)(n-1)},\;\;\;\;\;  m\neq n\label{eq:constraints}\\
X^9_{nn} & = & 2 \pi R_9 \identity +X^9_{(n-1)(n-1)}\ . \nonumber
\end{eqnarray}
Note that the matrix added to $X^9_{nn}$ is proportional to the
identity matrix.  This is because the translation operation only
shifts the diagonal components of the 0-brane matrices.  An easy way
to see this is that after $X^9$ has been diagonalized, its diagonal
elements correspond to the positions of the branes in direction
$X^9$; thus, adding a multiple of the identity matrix shifts the
positions by a constant amount.  Since the identity matrix commutes
with everything, this is the correct implementation of the translation
operation even when $X^9$ is not diagonal.

As a result of the constraints (\ref{eq:constraints}), the infinite block
matrix $X^9_{mn}$ can be written in the following form
\begin{equation}
\left(\begin{array}{ccccccc}
\ddots& X_{1} & X_{2} & X_{3} & \ddots \\
X_{-1} & X_{ 0} -2 \pi R_9\identity & X_{1} & X_{2} & X_{3} \\
X_{-2} & X_{-1}&X_0 & X_{1}&X_{2}  \\
X_{-3} &X_{ -2} & X_{-1} & X_{0}  + 2 \pi R_9\identity   &X_{1} \\
\ddots & X_{ -3} & X_{-2} & X_{ -1} & \ddots
\end{array} \right)
\label{eq:matrix-operator}
\end{equation}
where we have defined $X_k = X^9_{0k}$.

A matrix of this form can be interpreted as a matrix representation of
the operator
\begin{equation}
X^9 = i \hat{\partial} + A (\hat{x})
\end{equation}
describing the action of  a gauge covariant derivative
on a Fourier decomposition of functions of the form
\begin{equation}
\phi ( \hat{x}) = \sum_{n} \hat{\phi}_n e^{in \hat{x}/\hat{R}_9}
\end{equation}
which are periodic on a circle of radius $\hat{R}_9 = \alpha'/R_9 = 1/(2 \pi
R_9)$.
In order to see this correspondence concretely, let us first consider
the action of the derivative operator $i \hat{\partial}$ on such a
function.  Writing the Fourier components as a column vector
\begin{equation}
\phi (\hat{x}) \rightarrow \left(\begin{array}{c}
\vdots \\
\hat{\phi}_2 \\
\hat{\phi}_1 \\
\hat{\phi}_0 \\
\hat{\phi}_{-1} \\
\hat{\phi}_{-2} \\
\vdots
\end{array} \right)
\end{equation}
we find that the derivative operator acts as the matrix
\begin{equation}
i \hat{\partial}= {\rm diag}
(\ldots, -4 \pi R_9\identity, -2 \pi R_9\identity, 0,
2 \pi R_9\identity, 4 \pi R_9\identity, \ldots).
\end{equation}
This is precisely the inhomogeneous term along the diagonal of
(\ref{eq:matrix-operator})

Decomposing the connection $A (\hat{x})$ into Fourier components in turn
\begin{equation}
A (\hat{x}) =\sum_{n} A_n e^{in \hat{x}/\hat{R}_9}
\end{equation}
we find that multiplication of $\phi (\hat{x})$ by $A (\hat{x})$
precisely corresponds in the matrix language to the action of the
remaining part of (\ref{eq:matrix-operator}) on the column vector
representing $\phi (\hat{x})$, where $X_{n} = X_{0n}^9$ is identified
with $A_n$.

This shows that we can identify
\begin{equation}
X^9 \sim i  \hat{\partial}^9 + \hat{A}^9
\label{eq:relation1}
\end{equation}
under T-duality in the compact direction.  This identification
demonstrates that the infinite number of degrees of freedom in the
matrix $X^9$ of a constrained $U (\infty)$ Matrix theory describing
$N$ 0-branes on $\br^8 \times S^1$ can be precisely packaged in the
degrees of freedom of a $U(N)$ connection on a dual circle $\hat{S}^1$
of radius $\hat{R}_9 = 1/(2 \pi R_9)$.  A similar correspondence
exists for the transverse degrees of freedom $X^a$, $a < 9$, and for
the fermion fields $\psi$.  Because these fields are unchanged under
the translation symmetry, the infinite matrices which they are
described by in the 0-brane language satisfy condition
(\ref{eq:constraints}) without the inhomogeneous term.  Thus, these
degrees of freedom simply become $N \times N$ matrix fields
living on the dual $\hat{S}^1$ whose Fourier modes correspond to the winding
modes of the original 0-brane fields.

This construction gives a precise correspondence between the degrees
of freedom of the supersymmetric Matrix theory describing $N$ 0-branes
moving on $\br^8 \times S^1$ and the $(1 + 1)$-dimensional super
Yang-Mills theory on the dual circle.  To show that the theories
themselves are equivalent it only remains to check that the Lagrangian
of the 0-brane theory is taken to the super Yang-Mills Lagrangian
under this identification.  In fact, this is quite easy to verify.
Considering first the commutator terms, the term
\begin{equation}
{\rm Tr}\;[X^a, X^b]^2 \;\;\;\;\; (a, b \neq 9)
\end{equation}
in the 0-brane Matrix theory turns into the term
\begin{equation}
\frac{1}{2 \pi \hat{R}_9} \int d\hat{x} \; {\rm Tr}\;[X^a, X^b]^2
\end{equation}
of 2D super Yang-Mills.  Note that the trace in the 0-brane theory is
a trace over the infinite index $n \in\bz$ as well as over $i \in\{1,
\ldots, N\}$.  The trace over $n$ has the effect of extracting the
Fourier zero mode of the corresponding product of fields in the dual
theory.   The factor of $R_9 = 1/(2 \pi \hat{R}_9)$ in front of the integral in
the 2D super Yang-Mills is needed to normalize the zero mode so that
it integrates to unity.  Technically, there should be a factor of $1/|
\Gamma |$ multiplying the 0-brane matrix Lagrangian because of the
multiplicity of the copies; this factor is canceled by an overall
factor of $| \Gamma |$ from the trace, and since both factors are
infinite we simply drop them from all equations for convenience.  

Now let us consider the commutator term when one of the matrices is
$X^9$.  In this case we have
\begin{equation}
{\rm Tr}\;[X^9, X^a]^2
\end{equation}
which becomes after the replacement (\ref{eq:relation1})
\begin{equation}
-   \left(\frac{1}{2 \pi \hat{R}_9} \right)
\int d\hat{x} \; {\rm Tr}\; (\partial_9 X^a-i[A_9, X^a])^2
=-   \left(\frac{1}{2 \pi \hat{R}_9} \right)
\int d\hat{x} \; {\rm Tr}\;(D_9X^a)^2
\end{equation}
which is precisely the derivative squared term for the adjoint scalars
which we expect in the dual 1-brane theory.

The kinetic term for $X^9$ in the 0-brane theory becomes
the Yang-Mills curvature squared term in the dual theory
\begin{equation}
{\rm Tr}\; (D_0 X^9)^2 \rightarrow
  \left(\frac{1}{2 \pi \hat{R}_9} \right)
\int d\hat{x} \; {\rm Tr}\;F_{09}^2\ .
\end{equation}

The remaining terms in the 0-brane Lagrangian transform
straightforwardly into precisely the remaining terms expected in a 2D
super Yang-Mills Lagrangian with 16 supercharges.  This shows that
there is a rigorous equivalence between the low-energy field theory
description of $N$ 0-branes on $\br^8 \times S^1$ and the low-energy
field theory description of $N$ 1-branes wrapped around a dual $\hat{S}^1$
in the static gauge.  We note again the fact that the Lagrangian in
the dual Yang-Mills theory carries an overall multiplicative factor of
the original radius $R_9$.  This fact will play a significant role in
later discussions, particularly in regard to Matrix theory.
The fact that the coupling constant in the dual Lagrangian should
correspond with that of (\ref{eq:ym-coupling}) for a system of
1-branes indicates that under T-duality the string coupling transforms
through $\hat{g} = g \sqrt{\alpha'}/R_9$, which is what we expect from
string theory.

\vspace{0.08in}
\noindent
{\it \pbt\ \ \ $p$-branes on a torus $T^d$}
\vspace{0.05in}

So far we have discussed the situation of  $N$ 0-branes moving on a
space which has been compactified in a single direction.  It is
straightforward to generalize this argument to $p$-branes of arbitrary
dimension moving in a space with any number of compact dimensions.  By
carrying out the construction described above for each of the compact
directions in turn, it can be shown that the low-energy theory of $N$
$p$-branes which are completely unwrapped on a torus $T^d$ is
equivalent to the low-energy theory of $N$ $(p + d)$-branes which are
wrapped around the torus, in static gauge.  The only new type of term
which appears in the Lagrangian corresponds to a commutator term for
two directions which are both compactified.  In the original $p$-brane
theory, such a term would appear as $[X^a, X^b]^2$ (integrated over
the $p$-dimensional volume of the brane).
After T-duality on the two compact directions this term becomes
\begin{equation}
 -\left(\frac{1}{4 \pi^2 \hat{R}_a \hat{R}_b} \right)
\int d\hat{x}^a d\hat{x}^b \;  (F^{ab})^2
\end{equation}
which is just the appropriate Yang-Mills curvature strength squared
term in the dual theory.  Note that in the dual theory, the action is
multiplied by a factor of $R_a R_b$, since each compact direction
gives an extra factor of the radius.

As a particular example of compactification on a higher dimensional
torus, we can consider the theory of $N$ 0-branes on
a torus $T^d$.  After interpreting the winding modes of each matrix in
terms of Fourier modes of a dual theory, it follows that the
Lagrangian becomes precisely that of super Yang-Mills theory in $d +
1$ dimensions with a Yang-Mills coupling constant $g_{YM}$
proportional to $V^{-1/2}$ where $V$ is the volume of the original
torus $T^d$.

\vspace{0.08in}
\noindent
{\it \oat\ \ \ Further comments regarding T-duality}
\vspace{0.05in}

Throughout this section we have fixed the constant $2 \pi \alpha' =
1$.  It will be useful in some of the later discussions to have the
appropriate factors of $\alpha'$ reinstated in (\ref{eq:relation1}).
This is quite straightforward; since $\alpha'$ has units of length
squared, the correct T-duality relation is given by
\begin{equation}
X^a \leftrightarrow (2 \pi \alpha') (i \partial^a + A_a)
\label{eq:T-duality}
\end{equation}
where $X^a$ represents an infinite matrix of fields including winding
strings around a compactified dimension, and $A$ represents a
connection on a gauge bundle over the dual circle.

It should be emphasized that this T-duality relation
gives a precise correspondence between winding modes of strings on the
original circle and momentum modes on the dual circle.  This is
precisely the association  expected from
T-duality in perturbative string theory \mcite{dlp}.

So far we have been discussing field configurations in the $X^a$
matrices which correspond in the dual picture to connections on a
$U(N)$ bundle with trivial boundary conditions.  In fact, there are
also twisted sectors in the theory corresponding to bundles with
nontrivial boundary conditions.  We will now discuss such configurations
briefly.  To make the story clear, it is useful to reformulate the
above discussion in a slightly more abstract language.

The constraints (\ref{eq:constraints}) can be formulated by saying
that there exists a translation operator $U$ under which the infinite
matrices $X^a$  transform as
\begin{equation}
U X^a U^{-1} = X^a + \delta^{a9} 2 \pi R_9\identity.
\label{eq:constraint2}
\end{equation}
This relation is satisfied formally by the operators
\begin{equation}
X^9 = i \partial^9 + A_9, \;\;\;\;\; \;\;\;\;\;
U = e^{2 \pi i\hat{x}^9 R_9}
\end{equation}
which correspond to the solutions discussed above.  In this
formulation of the quotient theory, the operator $U$   generates the
group $\Gamma =\bz$ of covering space transformations.  Generally,
when we take a quotient theory of this type, however, there is a more
general constraint which can be satisfied.  Namely, the translation
operator only needs to preserve the state up to a gauge
transformation.  Thus, we can consider the more general constraint
\begin{equation}
U X^a U^{-1} = \Omega (X^a + \delta^{a9} 2 \pi R_9\identity) \Omega^{-1}.
\label{eq:generalconstraint}
\end{equation}
where $\Omega \in U(N)$ is an arbitrary element of the gauge group.
This relation is satisfied formally by
\begin{equation}
X^9 = i \partial^9 + A_9, \;\;\;\;\; \;\;\;\;\;
U = \Omega \cdot e^{2 \pi i\hat{x}^9 R_9} 
\end{equation}
This is precisely the same type of solution as we have above; however,
there is the additional feature that the translation operator now
includes a nontrivial gauge transformation.  On the dual circle
$\hat{S}^1$ this corresponds to a gauge theory on a bundle with a
nontrivial boundary condition in the compact direction 9.  Note that
even with such a nontrivial boundary condition, any $U(N)$ bundle over
$S^1$ is topologically trivial.
An example
of the type of boundary condition which might appear would be to take
$\Omega$ to be a permutation in $S_N$.  This type of gauge
transformation has the effect in the original 0-brane theory of
switching the labels of the 0-branes on each sheet of covering space.
When translated into the dual gauge theory picture, this corresponds
to a super Yang-Mills theory with a nontrivial boundary condition
$\Omega$ in the compact direction.

A similar story occurs when several directions are compact.  In this
case, however, there is a constraint on the translation operators in
the different compact directions.  For example, if we have
compactified on a 2-torus in dimensions 8 and 9, the generators $U_8$
and $U_9$ of a general twisted sector must generate a group isomorphic
to $\bz^2$ and therefore must commute.  The condition that these
generators commute can be related to the condition that the boundary
conditions in the dual gauge theory correspond to a well-defined
$U(N)$ bundle over the dual torus.  For compactifications in more than
one dimension such boundary conditions can define a topologically
nontrivial bundle.  In Section \ref{sec:bundles} we will discuss
nontrivial bundles of this nature in much more detail.  It is
interesting to note that this construction can even be generalized to
situations where the generators $U_i$ do not commute.  This leads to a
dual theory which is described by gauge theory on a noncommutative
torus \mcite{cds,Douglas-Hull,hww}.

%

{}
\subsection{S-duality for strings and super Yang-Mills}
\label{sec:S-duality}

The T-duality symmetry we have discussed above is a symmetry of type
II string theory which is essentially perturbative, in the sense that
the string coupling is only changed through multiplication by a
constant.  Another remarkable symmetry seems to exist in the type II
class of theories which is essentially nonperturbative; this is the
S-duality symmetry of the type IIB string
\mcite{Hull-Townsend,Schwarz-multiplet}.  S-duality is a symmetry which
acts according to the group $SL(2,{\bz})$ on the type IIB theory.  At
the level of the low-energy IIB supergravity theory, the dilaton and
axion form a fundamental $SL(2,\bz)$ multiplet, as do the NS-NS and
R-R two-forms.  Because the string coupling $g$ is given by the
exponential of the dilaton, this S-duality is a nonperturbative
symmetry which can exchange strong and weak couplings.  Because
symmetries in the S-duality group exchange the NS-NS and R-R
two-forms, we can see that S-duality exchanges strings and D1-branes,
and also exchanges D5-branes and NS (solitonic) 5-branes.  As there is
only a single four-form in the IIB theory, however, it must be left
invariant under S-duality; it follows that S-duality takes a D3-brane
into another D3-brane.

Since D3-branes are invariant under S-duality, it is interesting to ask
how we can understand the action of S-duality on the low-energy field
theory describing $N$ parallel D3-branes.  
This field theory is  the reduction to four dimensions of 
$U(N) \; {\cal N} = 1$ SYM in 10D, which is the pure $U(N)$
${\cal N} = 4$ super Yang-Mills
theory in 3 + 1 dimensions.  Since the Yang-Mills coupling of this
theory is related to the string coupling through $g_{YM}^2 \sim g$,
the action of S-duality on this super Yang-Mills theory must be a
nonperturbative $SL(2,\bz)$ duality symmetry.  In fact, for a number
of years it has been conjectured that 4D super Yang-Mills theory
with ${\cal N} = 4$ supersymmetry has precisely such an S-duality
symmetry.  This is a supersymmetric version of the non-abelian
S-duality symmetry proposed originally by Montonen and Olive.  We will
now briefly review the basics of this duality symmetry.  

Maxwell's equations describe a simple non-supersymmetric $U(1)$ gauge
theory in four dimensions.  In the absence of sources, these equations
have a very simple symmetry, which takes the curvature tensor $F$ to
its dual $*F$.  This has the effect of exchanging the electric and
magnetic fields in the theory (up to signs).  Although this symmetry
is broken when electric sources are introduced, if magnetic sources
are also introduced then the symmetry is maintained when the electric
and magnetic charges are also exchanged.

This marvelous symmetry of $U(1)$ gauge field theory seems at first
sight to break down for non-abelian theories with gauge groups like
$U(N)$.  It was suggested, however, by Montonen and
Olive \mcite{Montonen-Olive} that such a symmetry might be possible for
non-abelian theories if the gauge group $G$ were replaced by a dual
group $\hat{G}$ with a dual weight lattice.  Further
work \mcite{Witten-Olive,Osborn} indicated that such a non-abelian
duality symmetry would probably only be possible in theories with
supersymmetry, and that the ${\cal N} = 4$ theory was the most likely
candidate.  Although there is still no complete proof that the ${\cal
N} = 4$ super Yang-Mills theory in 4D has this S-duality symmetry,
there is a growing body of evidence which supports this conclusion.

The proposed non-abelian S-duality symmetry of 4D super Yang-Mills
acts by the group $SL(2,\bz)$, just as we would expect from string
theory.  The (rescaled) Yang-Mills coupling constant and theta angle
can be conveniently packaged into the quantity
\begin{equation}
\tau = \frac{\theta}{2 \pi}  + \frac{i}{g_{{\rm YM}}^2} 
\end{equation}
which is transformed under
$SL(2,{\bz})$  by the standard transformation law
\begin{equation}
\tau \rightarrow \frac{a \tau + b}{c \tau + d} 
\end{equation}
where $a, b, c, d \in {\bz}$ with $ad-bc = 1$ parameterize a matrix
\begin{equation}
\left(\begin{array}{cc}
a & b\\c & d
\end{array}\right)
\end{equation}
in $SL(2,\bz)$.
In particular, the group $SL(2,\bz)$ is generated by the transformations
\begin{equation}
\tau \rightarrow \tau + 1
\end{equation}
corresponding to the periodicity of $\theta$, and
\begin{equation}
\tau \rightarrow -1/\tau
\end{equation}
which inverts the coupling and corresponds to strong-weak duality.

There is by now a large body of evidence that S-duality is a true
symmetry of ${\cal N} = 4$ super Yang-Mills.  However, to date there
is no real proof of S-duality from the point of view of field theory.
One of the strongest pieces of evidence for this duality symmetry is
the fact that the spectrum of supersymmetric bound states of dyons is
invariant under the action of the S-duality group; a detailed proof of
this result and further references can be found in \mycite{Ferrari}.


\section{Branes and Bundles}
\label{sec:bundles}

As we discussed in Section \oat, there are different topological
sectors for a system of 0-branes on a torus which correspond in the
dual gauge theory language to nontrivial $U(N)$ bundles over the dual
torus.  In fact, these topologically nontrivial configurations of
branes correspond to systems containing not only the original 0-branes
but also branes of higher dimension.  In this section we describe in
some detail a general feature of D-branes which amounts to the fact
that the low-energy Yang-Mills theory describing Dirichlet $p$-branes
also contains information about D-branes of both higher and lower
dimensions.  Roughly speaking, D-branes of lower dimension can be
described by topologically nontrivial configurations of the $U(N)$
gauge field living on the original $p$-branes, while D-branes of
higher dimension can be encoded in nontrivial commutation relations
between the matrices $X^a$ describing transverse D-brane excitations
in compact directions.  In order to make the discussion precise, it
will be useful to begin with a review of nontrivial gauge bundles on
compact manifolds.  In these notes we will concentrate primarily on
configurations of D-branes on tori; on general compact spaces the
story is similar but there are some additional subtleties
\mcite{bvs,ghm,Cheung-Yin,Minasian-Moore}.

\subsection{Review of vector bundles}

An introductory review of bundles and their relevance for gauge field
theory is given in \mycite{egh}.  In this section we briefly review some
salient features of bundles and Yang-Mills connections.
Roughly speaking, a (real) vector bundle is a space constructed by
gluing together a copy of a vector space $V =\br^k$ (called the fiber
space) for each point on a particular manifold $M$ (called the base
manifold) in a smooth fashion.  Mathematically speaking, a vector
bundle can be defined by decomposing $M$ into coordinate patches
$U_i$.  The vector bundle is locally equivalent to $U_i \times
\br^k$.  When the patches of $M$ are glued together, however, there can
be nontrivial identifications which give the vector bundle a
nontrivial topology.  For every pair of intersecting patches $U,   V$
there is a transition function between these patches which relates the
fibers at each common point.  Such a transition function $\Omega_{UV}$
takes values in a group $G$ called the structure group of the bundle.
The transition function $\Omega_{UV}$ identifies
$(u,f')$ and $(v, f)$ where $u$ and $v$ are points in $U$ and $V$
which represent the same point in $M$, and where the fiber elements
are related through $f' =\Omega_{UV}f$.

\begin{figure}
\vspace{-0.3in}
\psfig{figure=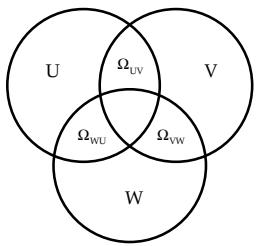,height=1.8in}
\vspace{-0.15in}
\caption[x]{\footnotesize Three patches on a manifold $M$ over which a
bundle is defined}
\label{f:cocycle}
\end{figure}

In order to describe a well-defined bundle, the transition functions
must obey certain relations called cocycle conditions.  For example,
if as in Figure~\ref{f:cocycle} there are three patches $U, V, W$
whose intersection is nonempty, the transition functions between
the three patches must obey the relation
\begin{equation}
\Omega_{UV} \Omega_{VW} \Omega_{WU} = {\rm id}
\end{equation}
where ${\rm id}$ is the identity element in $G$.
This is clearly necessary in order that a point $(u, f)$ in the
intersection region not be identified with any other point in the same
fiber after repeated application of the transition functions.

This describes a bundle whose fiber is a vector space.  Another type
of bundle, called a principal bundle, has a fiber which is a copy of
the structure group $G$ itself.
A Yang-Mills connection for a gauge theory with gauge group $G$
is associated with a principal bundle with
fiber $G$.  Formally speaking, a Yang-Mills connection $A_\mu$ is a
one-form which takes values in the Lie algebra of $G$.  A connection
of this type gives a definition of parallel transport in the bundle.
The most
important feature for our purposes is the transformation property of
such a connection under a transition function $\Omega$, which is given by
\begin{equation}
A' = \Omega \cdot A \cdot \Omega^{-1} -i\; d\Omega \cdot \Omega^{-1}
\end{equation}

Generally,  a physical theory will include both a Yang-Mills field and
additional matter fields.  The Yang-Mills connection is defined with
respect to a particular principal bundle, and the matter fields are
given by sections of associated vector bundles whose transition
functions are given by particular representations of the $G$-valued
transition functions of the principal bundle.

Over any compact Euclidean manifold, such as the torus $T^{d}$, there
are many topologically inequivalent ways to construct a nontrivial
bundle.  One way to distinguish such inequivalent bundles is through
the use of topological invariants called characteristic classes.  One
of the simplest examples of characteristic classes are the Chern
classes.  These classes distinguish topologically inequivalent
$U(N)$ bundles, and are given by invariant polynomials in the Yang-Mills
field strength
\begin{equation}
F_{\mu \nu} = \partial_\mu A_\nu -\partial_\nu A_\mu -i[A_\mu, A_\nu].
\end{equation}
The first two Chern classes are defined by
\begin{eqnarray}
c_1 & = &\frac{1}{2 \pi} {\rm Tr}\; F\\
c_2 & =  &\frac{1}{8 \pi^2} \left(
{\rm Tr}\; F \wedge F-({\rm Tr}\; F) \wedge({\rm Tr}\; F) \right)  \nonumber
\end{eqnarray}
These forms $c_i$ are integral cohomology classes, so that when $c_i$
is integrated over any $2i$-dimensional submanifold (homology class)
the result is an integer.  As we will now discuss, in a low-energy
D-brane Yang-Mills theory, these integers count lower-dimensional
D-branes embedded in the original D-brane world-volume.

\subsection{D-branes from Yang-Mills curvature}
\label{sec:branescurvature}

Let us consider the low-energy Yang-Mills theory describing $N$ coincident
$p$-branes.  If the bundle associated with the Yang-Mills connection
is nontrivial, this indicates that the gauge field configuration
carries R-R charges which are associated with D-branes of
dimension less than $p$.  We will now formulate precisely the way in
which these lower-dimensional D-branes appear, after which we will
discuss the justification for these statements.

Simply put, the integral form corresponding to the $k$th antisymmetric
product of the curvature form $F$ carries $(p-2k)$-brane charge.
Thus,
%
\begin{eqnarray}
&\makebox[1in][r]{$\frac{1}{2 \pi} \int {\rm Tr}\; F$}  & {\rm
corresponds\ to} \; 
(p-2)-{\rm brane\ charge}.\nonumber\\ 
&\makebox[1in][r]{$\frac{1}{8 \pi^2} \int {\rm Tr}\;  (F \wedge F)$}
& {\rm corresponds\ to}
 \; (p-4)-{\rm brane\ charge}. \nonumber\\ 
&\makebox[1in][r]{$\frac{1}{48 \pi^3}
 \int {\rm Tr}\; (F \wedge F \wedge F)$}  & {\rm corresponds\ to} \;
(p-6)-{\rm brane\ charge,\ etc.} \ldots  \nonumber
\end{eqnarray}
More precisely, let us imagine that a $(p-2)$-brane is wrapped around
some $(p-2)$-dimensional homology cycle $h_{p-2}$ in the
$p$-dimensional volume of the original $p$-branes.  If we choose any
two-dimensional cycle $h_2$, it will generically intersect $h_{p-2}$
in a fixed number of points, corresponding to the intersection number
of these two cycles.  Thus, any $(p-2)$-cycle defines a cohomology
class which associates an integer with any 2-cycle.  This cohomology
class is known as the Poincare dual of the original homology class.
Using this correspondence we can state the connection between D-branes
and field strength precisely: The integral cohomology class
proportional to $F \wedge^k F$ is the Poincare dual of a
$(p-2k)$-dimensional homology class which describes a system of
embedded $(p-2k)$-branes.

The observation that the instanton number $\frac{1}{8 \pi^2} \int F
\wedge F$ carries 
$(p-4)$-brane charge was first made by Witten in the context of
5-branes and 9-branes \mcite{Witten-small}.  The more general result
for arbitrary $p, k$ was described by Douglas \mcite{Douglas}.  
From the string theory point of view, this correspondence between
D-branes and Yang-Mills curvature arises from  a Chern-Simons type of
term which appears in the full D-brane action, and is given by \mcite{Li-bound}
\begin{equation}
{\rm Tr}\; \int_{\Sigma_{p + 1}} {\cal A} \wedge e^{F}
\label{eq:Chern-Simons}
\end{equation}
where ${\cal A}$ is a sum over all the R-R fields
\begin{equation}
{\cal A} = \sum_{k} A^{(k)}
\end{equation}
and where the integral is taken over the full $(p + 1)$-dimensional
world-volume of the $p$-brane.
For example, on a 4-brane $F \wedge F$ couples to $A^{(1)}$ through
\begin{equation}
{\rm Tr}\;\int_{\Sigma_{5}}A^{(1)} \wedge F \wedge F,
\end{equation}
demonstrating that $F \wedge F$ is playing the role of 0-brane charge
in this case.
The existence of the Chern-Simons term (\ref{eq:Chern-Simons}) can be
shown on the basis of anomaly cancellation arguments \mcite{ghm}.
It is also possible, however, to show that
these terms must appear simply using the principles of T-duality
and rotational invariance
\mcite{Bachas,Douglas,bdgpt,abb}.  We  follow the latter approach here; in
the following sections we  show how the correspondence between
lower-dimensional branes and wedge products of the curvature form can
be seen directly in the low-energy Yang-Mills description of D-branes,
using only T-duality and the intrinsic properties of Yang-Mills theory.

\subsection{Bundles over tori}

We will be primarily concerned here with D-branes on
toroidally compactified spaces.  Thus, it will be useful to explicitly
review here some of the properties of $U(N)$ bundles over tori.  Let
us begin with the simplest case, the two-torus $T^2$.

\begin{figure}
\vspace{-0.3in}
\psfig{figure=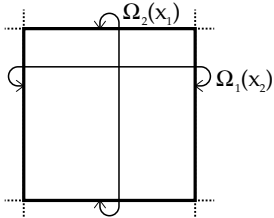,height=1.5in}
\vspace{-0.15in}
\caption[x]{\footnotesize Transition
functions defining a bundle over the 2-torus}
\label{f:torus}
\end{figure}

If we consider a space which has been compactified on $T^2$ with radii
\begin{equation}
R_1 = L_1/(2 \pi), \; R_2 = L_2/(2 \pi)
\end{equation}
then the low-energy field theory of $N$ wrapped 2-branes is $U(N)$ SYM
on $T^2$.  To describe a $U(N)$ bundle over a general manifold, we
need to choose a set of coordinate patches on the manifold.  For the
torus, we can choose a single coordinate patch covering the entire
space, where the transition functions for the bundle are given by (see
Figure~\ref{f:torus})
\begin{equation}
\Omega_1 (x_2), \;\Omega_2 (x_1).
\end{equation}
A connection on a bundle defined by these transition functions must
obey the boundary conditions
\begin{eqnarray}
A_1  (x_1 + L_1, x_2) & = &  \Omega_1 (x_2) A_1 (x_1, x_2) \Omega_1^{-1}
(x_2) \\
A_2  (x_1 + L_1, x_2) & = &  \Omega_1 (x_2) A_2 (x_1, x_2) \Omega_1^{-1}
(x_2)-i \; (\partial_2 \Omega_1 (x_2)) \cdot \Omega^{-1}_1 (x_2) \nonumber\\
A_1  (x_1, x_2 + L_2) & = &  \Omega_2 (x_1) A_1 (x_1, x_2) \Omega_2^{-1}
(x_1)-i \; (\partial_1 \Omega_2 (x_1)) \cdot \Omega^{-1}_2 (x_1)\nonumber\\
A_2  (x_1, x_2 + L_2) & = &  \Omega_2 (x_1) A_2 (x_1, x_2) \Omega_2^{-1}
(x_1)\nonumber
\end{eqnarray}
while a matter field $\phi$ in the fundamental representation must
satisfy the boundary conditions
\begin{eqnarray}
\phi (x_1 + L_1, x_2) & = & \Omega_1 (x_2)\phi (x_1, x_2)  \nonumber\\
\phi (x_1, x_2 + L_2) & = & \Omega_2 (x_1)\phi (x_1, x_2).
\end{eqnarray}
The cocycle condition for a well-defined $U(N)$ bundle is
\begin{equation}
 \Omega_1 (L_2) \Omega_2 (0)  \Omega_1^{-1} (0)  \Omega_2^{-1} (L_1)=
 \identity.
\label{eq:bc}
\end{equation}

In general, $U(N)$ bundles over $T^2$ are classified by the first
Chern number 
\begin{equation}
C_1 = \int c_1=\frac{1}{2 \pi} \int {\rm Tr}\; F = k \in {\bz}
\end{equation}
Physically, this integer corresponds to the total non-abelian magnetic
flux on the torus.  In order to understand these nontrivial $U(N)$
bundles, it is helpful to decompose the gauge group into its abelian
and non-abelian components
\begin{equation}
U(N) = (U(1) \times SU(N))/{\bz}_N.
\label{eq:un}
\end{equation}
Because the curvature $F$ has a trace which arises purely from the
abelian part of the gauge group, we see that the $U(1)$ part of the
total field strength for a bundle with $C_1 = k$ is given by $F =
k\identity/N$.  Such a field strength would not be possible for a purely
abelian theory (assuming the existence of matter fields in the
fundamental representation) since it would not be possible to satisfy
(\ref{eq:bc}).  Once $U(1)$ is embedded in $U(N)$ through
(\ref{eq:un}), however, this deficiency can be corrected by choosing $SU(N)$
boundary conditions $\tilde{\Omega}$ which correspond to a ``twisted''
bundle.  Such boundary conditions satisfy
\begin{equation}
 \tilde{\Omega}_1 (L_2) \tilde{\Omega}_2 (0)
  \tilde{\Omega}_1^{-1} (0)  \tilde{\Omega}_2^{-1} (L_1)=Z
\end{equation}
where $Z = e^{-2 \pi ik/N} \identity$ is central in $SU(N)$.  Twisted bundles of this
type were originally considered by 't Hooft \mcite{Hooft}.  

The integer $k$ gives a complete classification of $U(N)$ bundles over
$T^2$.  Over a higher dimensional torus $T^n$, the story is
essentially the same, however there is an integer $k_{ij}$ for every
pair of dimensions in the torus.  For each dimension $i$ there is a
transition function, and for each pair $i, j$ the transition functions
satisfy a cocycle relation of the form (\ref{eq:bc}).

\subsection{Example: 0-branes as flux on $T^2$}
\label{sec:examplet2}

We will now discuss nontrivial bundles on $T^2$ and show
using T-duality that the first Chern class indeed counts 0-branes.

Consider a $U(N)$ theory on $T^2$ with total flux $\int {\rm Tr}\; F = 2 \pi$.
We can choose an explicit set of boundary conditions corresponding to
such a bundle
\begin{eqnarray}
\Omega_1 (x_2) & = & e^{2 \pi i (x_2/L_2) T}V \label{eq:boundary2}\\
\Omega_2 (x_1) & = & \identity \nonumber
\end{eqnarray}
where
\begin{equation}
V =
\pmatrix{
 & 1 & && \cr
 &   & 1 && \cr
 &   &   & \ddots & \cr
& & & & 1\cr
1&   &   &  & 
} 
\end{equation}
and $T = {\rm Diag} \; (0, 0, \ldots, 0, 1)$.

\begin{figure}
\psfig{figure=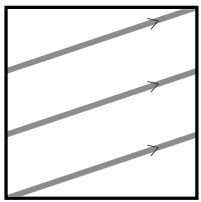,height=1.5in}
\caption[x]{\footnotesize An $(N, 1)$ diagonally wrapped string dual to
$N$ 2-branes with one unit of flux}
\label{f:diagonal}
\end{figure}

To understand the D-brane geometry of this bundle, let us construct a
linear connection on the bundle, which will correspond in the T-dual
picture to flat D-branes on the dual torus.  The boundary conditions
(\ref{eq:boundary2}) admit a linear connection with constant curvature
\begin{equation}
\begin{array}{lll}
A_1 & = & 0  \\
A_2 & = & F x_1 + \frac{2 \pi}{L_2}  \;
{\rm Diag} (0,  1/N, \ldots, (N -1)/N)
\end{array}
\end{equation}
with
\begin{equation}
F = \frac{2 \pi}{N L_1 L_2} \identity.
\end{equation}
Because we have chosen the boundary conditions such that $\Omega_2 =
1$ we can T-dualize in a straightforward fashion using $X^2 = (2 \pi
\alpha') (i
\partial_2 + A_2)$.
After such a T-duality,  $X^2$ represents the transverse positions of
a set of
1-branes on $T^2$.  This field is represented by an infinite matrix
with indices $n \in \bz$ and $i \in\{1, \ldots, N\}$.  In the $(n, m)
= (0, 0)$ block, the field $X^2$ is given by the matrix 
\begin{equation}
X^2 = 
 \left( \frac{\hat{L}_2}{N}  \right) 
\left(\begin{array}{cccc}
\frac{x_1}{L_1} & 0 & 0 & \ddots\\
0 &\frac{x_1}{L_1} + 1 & 0 & \ddots\\
0 &0 & \ddots & 0\\
\ddots &  \ddots & 0 &\frac{x_1}{L_1} + (N -1)
\end{array}\right)
\end{equation}
where
\begin{equation}
\hat{L}_2 =\frac{4 \pi^2 \alpha'}{L_2}
\end{equation}
Thus, we see that the T-dual of the original gauge field on $T^2$
describes a single 1-brane wrapped once diagonally around $\hat{R}_2$, and
$N$ times around $R_1$ (See Figure~\ref{f:diagonal}).  

The dual configuration has quantum numbers corresponding to
$N$ 1-branes on $R_1$ and a single 1-brane on $\hat{R}_2$.  In homology
this state could be written as
\begin{equation}
N \cdot (1) + (2)
\end{equation}
Since wrapped 1-branes are T-dual to 0-branes, the original flux on
the 2-brane corresponds to a single 0-brane.  This gives a simple
geometrical demonstration through T-duality of the result that the first
Chern class counts $(p-2)$-branes.  

It is straightforward to carry out
an analogous construction for a system with $k$ 0-branes.  In this
case, the nontrivial boundary condition becomes \mcite{ht}
\begin{equation}
\Omega_1 (x_2)  =  e^{2 \pi i (x_2/L_2) T}V^k
\end{equation}
with $T$  being the diagonal matrix
\begin{eqnarray}
T
 & = &  {\rm Diag} (n, \ldots, n, n + 1, \ldots, n + 1)
\end{eqnarray}
where $n$ is the integral part of $k/N$
and where the multiplicities
of the diagonal elements of $T$ are $N-k$ and $k$ respectively.  

In the discussion in this section we have chosen to set $\Omega_2 = 1$
for convenience.  This makes the discussion slightly simpler since the
T-duality relation in direction 2 is implemented directly through
(\ref{eq:T-duality}).  For a nontrivial gauge transformation
$\Omega_2$  T-duality would give a
configuration of the type described by
(\ref{eq:generalconstraint}).  For example, if we used the more
standard ('t Hooft type) boundary conditions for the bundle with $k = 1$
\begin{eqnarray}
\Omega_1 (x_2) & = &  e^{2 \pi i (x_2/ L_2) (1/N)}U\\
\Omega_2 (x_1) & = &  V\nonumber
\end{eqnarray}
where
\begin{equation}
U = 
\pmatrix{
1& & & \cr
& e^{2\pi i\over N} & & \cr
& & \ddots & \cr
& & & e^{2\pi i (N-1)\over N}
} 
\end{equation}
Then after T-duality in direction 2 we would get a 1-brane
configuration in which translation by $2 \pi \hat{R}_2$ in the
covering space would give rotation by $V$, permuting the labels on the
1-branes.  This situation is gauge equivalent to the one we have
discussed where the boundary conditions are given by
(\ref{eq:boundary2}).

\subsection{Example: 0-branes as instantons on $T^4$}
\label{sec:examplet4}

Let us now consider nontrivial bundles on $T^4$.  From the previous
discussion it is clear that a nonvanishing first Chern class indicates
the existence of 2-branes in the system.  For example, if $\int F_{12}
= 2 \pi$ then the configuration contains a 2-brane wrapped around the
(34) homology cycle.  A constant curvature connection with $\int
F_{12} = 2 \pi$ and $\int F_{34} = 2 \pi$ would correspond after
T-duality in directions 2 and 4 to a diagonally wrapped 2-brane, and
in the original Yang-Mills theory on $T^4$ corresponds to a
``4-2-2-0'' configuration with a unit of 0-brane charge as well as
2-brane charge in directions $(12)$ and $(34)$ \mcite{Sanjaye-Zack}.  A
more interesting configuration to consider is one where the first
Chern class vanishes but the second Chern class does not.  This
corresponds to an instanton in the $U(N)$ gauge theory on $T^4$.  To
consider an explicit example of such a configuration, let us take a
$U(N)$ gauge theory on a torus $T^4$ with sides all of length $L$.  We
want to construct a bundle with nontrivial second Chern class $C_2 =
(8 \pi^2)^{-1} \int c_2 = k$ and with $c_1 = 0$.  There is no smooth
$U(N)$ instanton with $k = 1$; a single instanton tends to
shrink to a point on the torus \mcite{vb-shrink}.  Thus, we will
consider a configuration with $k = 2, N = 2$.

To construct a bundle with the desired topology we can take the
transition functions in the four directions of the torus to be
\begin{eqnarray}
\Omega_2 = \Omega_4 & = & \identity \nonumber\\
\Omega_1 & = &  e^{2 \pi i (x_2/L) \tau_3}\\
\Omega_3 & = &  e^{2 \pi i (x_4/L) \tau_3} \nonumber
\end{eqnarray}
where
\begin{equation}
\tau_3 =\left(\begin{array}{cc}
1 & 0\\
0 & -1
\end{array} \right)
\end{equation}
is the usual Pauli matrix.
This bundle admits a linear connection 
\begin{eqnarray}
A_1 = A_3  & = &  0 \nonumber\\
A_2 & = &  \frac{2 \pi x_1}{L^2} \tau_3\\
A_4 & = &  \frac{2 \pi x_3}{L^2}  \tau_3 \nonumber
\end{eqnarray}
whose curvature is given by
\begin{eqnarray}
F_{12} = F_{34}  & = & \frac{2 \pi}{L^2}  \tau_3
\end{eqnarray}
Since ${\rm Tr}\; F = 0$ there is no net 2-brane charge, as desired.
The instanton number of the bundle is
\begin{equation}
C_2 = \frac{1}{8 \pi^2}  \int d^4 x \; {\rm Tr}\;F \wedge F = 2.
\end{equation}
As we would expect from the discussion in section
\ref{sec:branescurvature}, this should correspond to the existence of
two 0-branes in the system.  We can see this by again using T-duality.
After performing T-duality transformations in directions 2 and 4 we get
two 2-branes whose transverse coordinates are described by  
\begin{eqnarray}
X^2 (x_1, x_3) & = &  \pm\hat{L}x_1/L\\
X^4 (x_1, x_3) & = &  \pm \hat{L} x_3/L\nonumber
\end{eqnarray}
where $\hat{L} =4 \pi^2 \alpha'/L$.
These 2-branes are wrapped diagonally on the dual $T^4$ in such a way
that they correspond to the following homology 2-cycles
\begin{eqnarray}
{\rm brane\ 1} & \rightarrow & 
(13) + (14) + (23) + (24)\\
{\rm brane\ 2} & \rightarrow & 
(13) - (14) - (23) + (24) \nonumber
\end{eqnarray}
The total resulting homology class is $2(13)+ 2 (24)$, which is
T-dual to two 4-branes and two 0-branes as expected.  Further
discussion of configurations of this type which are dual to instantons
on $T^4$ can be found in \mycite{bdl,Sanjaye-Zack,ht}.

{}
\subsection{Branes from lower-dimensional branes}
\label{sec:branes-smaller}

In the preceding subsections we have discussed how, in general,
$(p-2k)$-branes can be described by nontrivial gauge configurations in
the world-volume of a system of parallel $p$-branes.  We will now
discuss the T-dual interpretation of this result, which indicates that
it is equally possible to construct $(p + 2k)$-branes out of a system
of interacting $p$-branes by choosing noncommuting matrices to
describe the transverse coordinates.

In the context of the preceding discussion, it is easiest to describe
the construction of higher-dimensional branes from a finite number of
$p$-branes in the case of toroidally compactified space.  In Sections
\mom\ and \molf\ we will discuss the construction of higher dimensional
branes in noncompact spaces from a system of 0-branes.  The simplest
example of the phenomenon we wish to discuss here is the description
of a 2-brane in terms of a ``topological'' charge associated with the
matrices describing $N$ 0-branes on $T^2$.  To see how a configuration
with such a charge is constructed, consider again the diagonal ($N,
1$) 1-brane on $T^2$ (Figure~\ref{f:diagonal}).  If we take the
toroidal dimensions to be $L_1 \times L_2$ then the diagonal 1-brane
configuration satisfies
\begin{equation}
[(\partial_1 -iA_1), X^2] = \frac{L_2}{N L_1}  \identity.
\end{equation}
By taking the
T-dual on $X^2$  we get a system of $N$ 2-branes with unit flux
\begin{equation}
[(\partial_1-iA_1), (\partial_2-iA_2)] = -iF =
\frac{-2 \pi i}{N L_1 \hat{L}_2}  \identity,
\end{equation}
as discussed in Section \ref{sec:examplet2}.  If, on the other hand,
we perform a T-duality transformation on $X^1$, then we get a system
of $N$ 0-branes satisfying
\begin{equation}
[X^1, X^2] = \frac{2 \pi \hat{R}_1 R_2 i}{N} \identity
\end{equation}
where $\hat{R}_1$ and $R_2$ are the radii of the torus on which the
0-branes are moving.  Since the 1-brane wrapped around $X^2$ becomes a
2-brane on $T^2$ under the T-duality transformation which takes the
1-branes on $X^1$ to 0-branes, we see that on a $T^2$ with area $A$ a
system of $N$ 0-branes described by (infinite) matrices $X$ satisfying
\begin{equation}
{\rm Tr}\; [X^1, X^2] = \frac{iA}{2 \pi} 
\label{eq:2-branecharge}
\end{equation}
carries a unit of 2-brane charge.
Note that if the 0-branes were not moving on a compact space the
quantity in (\ref{eq:2-branecharge}) would vanish for $N$ finite.
In the infinite $N$ limit, however, as will be discussed in
\mom, this charge can be nonzero even in Euclidean space.

This discussion generalizes naturally to higher dimensions.  For
example, a system of $N$ 0-branes on a $T^4$ of volume $V$ with
\begin{equation}
{\rm Tr}\;\epsilon_{abcd} X^a X^b X^c X^d = \frac{V}{2 \pi^2} 
\end{equation}
will carry a unit of 4-brane charge \mcite{grt,bss}.  This is just the
T-dual of the instanton number for a system of $N$ 4-branes, which is
associated with 0-brane charge as discussed above.  Similarly, any
system of $p$-branes on a $2k$-dimensional transverse torus can be in
a state with $(p+2k)$-brane charge.

It is also, of course, possible to mix the two types of conditions we
have discussed to describe, for example, 2-brane charge on the (34)
homology cycle of a 4-torus in terms of a gauge theory of 2-branes
wrapped on the (12) homology cycles.  Such a charge is proportional to
\begin{equation}
{\rm Tr}\; \left(F_{12}[X^3, X^4] -(D_1 X^3) (D_2 X^4)
+ (D_1 X^4) (D_2 X^3) \right).
\end{equation}

\subsection{Strings and electric fields}
\label{sec:strings-electric}

We have seen that the gauge fields and transverse coordinates of a system of
$p$-branes can be combined to give $(p \pm 2k)$-brane charge.  It is
also possible to choose gauge fields on the world-volume of a
$p$-brane which describe fundamental strings.  
Consider a system of 0-branes moving on  a space which has been
compactified in direction $X^9$.
Clearly, these 0-branes can be given momentum in the compact
direction; this momentum is proportional to $\dot{X}^9$ and is
quantized in units of $1/R$.
Under T-duality on the $S^1$, we have
\begin{equation}
\dot{X^9} \rightarrow \int (2 \pi \alpha')\dot{A_9}
\end{equation}
Thus, momentum of a set of 0-branes corresponds to electric flux
around the compact direction in the dual gauge theory.  Since string
momentum is T-dual to string winding, we see that electric flux in a
gauge theory on a compact space can be associated with fundamental
string winding number.  It is natural to give this result a local
interpretation, so that lines of electric flux in a gauge theory
correspond to fundamental strings even in noncompact space.

It is interesting to note that  0-brane momentum in
a compact direction and the T-dual string winding number are quantized
only because of the quantum nature of the theory.  On the other hand,
the quantization of flux giving 0-brane charge in a gauge theory on
$T^2$ arises from topological considerations, namely the fact that the
first Chern class of a $U(1)$ bundle is necessarily integral.
Nonetheless, in string theory these quantities which are quantized in
such different fashions can be related through duality.
It is tempting to speculate that a truly fundamental description of
string theory would therefore in some way combine quantum mechanics
and topological considerations in a novel fashion.

{}

\normalsize

\section{D-brane Interactions}



So far we have discussed the geometry of D-branes as described by
super Yang-Mills theory.  We now proceed to describe some aspects of
D-brane interactions.  We begin with a discussion of D-brane bound
states from the point of view of Yang-Mills theory.  We then discuss
potentials describing interactions between separated D-branes.

\subsection{D-brane bound states}
\label{sec:bound}

Bound states of D-branes were originally understood from supergravity
(as discussed in the lectures of Stelle at this school \mcite{Stelle})
and by duality from the perturbative string spectrum
\mcite{Hull-Townsend,Sen-bound1,Sen-bound2}.  There are a number of
distinct types of bound states which are of interest.  These include
$p-p'$ bound states between D-branes of different dimension, $p-p$
bound states between identical D-branes, and bound states of D-branes
with strings.  We will discuss each of these systems briefly; in order
to motivate the results on bound states, however, it is now useful to
briefly review the concept of BPS states.

\vspace{0.08in}
\noindent
{\it \bps\ \ \ BPS states}
\vspace{0.05in}

Certain extended supersymmetry (SUSY) algebras contain central terms,
so that the full SUSY algebra has the general form
\begin{equation}
\{Q, Q\}  \sim P + Z.
\end{equation}
For example, in $D = 4$, ${\cal N} = 2$ $U(2)$ super
Yang-Mills \mcite{Witten-Olive},
\begin{equation}
\{Q_{\alpha i}, \bar{Q}_{\beta j}\} =
\delta_{ij} \gamma^{\mu}_{\alpha \beta} P_\mu
+ \epsilon_{ij} \left(\delta_{\alpha \beta} U +
(\gamma_5)_{\alpha \beta} V\right).
\end{equation}
where
\begin{equation}
U =\langle \phi \rangle e \;\;\;\;\; V = \langle \phi \rangle g
\end{equation}
are related to electric and magnetic charges after spontaneous
breaking to $U(1)$.
Since $\{Q_{\alpha i},  \bar{Q}_{\beta j}\}$ is positive definite it follows
that
\begin{equation}
M^2 \geq U^2 + V^2
\end{equation}
so
\begin{equation}
M \geq \langle \phi \rangle \sqrt{e^2 + g^2}
\label{eq:BPS-inequality}
\end{equation}

This inequality is saturated when $\{Q_{\alpha i}, \bar{Q}_{\beta
j}\}$ has vanishing eigenvalues.  This condition implies $Q | {\rm
state} \rangle = 0$ for some $Q$.  Thus, any state with a mass
saturating the inequality (\ref{eq:BPS-inequality}) lies in a
``short'' representation of the supersymmetry algebra.  Because this
property is protected by supersymmetry, it follows that the relation
between the mass and charges of such a state cannot be modified by
perturbative or nonperturbative effects (although the mass and charges
can be simultaneously modified by quantum effects).

{}

Similar BPS states appear in string theory, where the central terms in
the SUSY algebra correspond to NS-NS and
R-R charges.  As in the above example,
states which preserve some SUSY are BPS saturated.  There are many
ways of analyzing BPS states in string theory.  The spectrum of BPS
states with a particular set of D-brane charges can in some cases be
determined through duality from perturbative string states
\mcite{Hull-Townsend,Sen-bound1,Sen-bound2}.
Such dualities allow the number of BPS states with fixed charges to be
counted.  BPS states can also be found through the space-time
supersymmetry algebra \mcite{Polchinski-TASI}, providing a
connection to the large body of known results on supergravity
solutions \mcite{Stelle}.  We can also analyze BPS states using the
Yang-Mills or Born-Infeld theory on the world-volume of a set of
D-branes.  We will follow this latter approach in the next few sections.

Before discussing BPS bound states in detail, let us synopsize results
on the energies of these states which can be obtained from duality or
the supersymmetry algebra \mcite{Polchinski-TASI}.  We will then show
that these results are correctly reproduced in the SYM description.

\vspace{0.05in}\noindent
{\sl i}. $p-p$ BPS systems are marginally bound.  This means that the
energy of a bound state of $N$ $p$-branes, when such a state exists,
is $N E_p$ where $E_p$ is the 
energy of a single $p$-brane.

\vspace{0.03in}\noindent
{\sl ii}. $p-(p+4)$ BPS systems are marginally bound.  For a bound
state of $N_p$ $p$-branes and $N_{p +4}$ $(p+4)$-branes the total
energy is  $E = N_pE_p + N_{p + 4} E_{p +4}$.

\vspace{0.03in}\noindent
{\sl iii}. $p-(p+2)$ BPS systems are truly bound when $N_p$ and $N_{p
+2}$ are relatively prime.  For
these systems, the energy is $E = \sqrt{(N_pE_p)^2 + (N_{p +2}E_{p +2})^2}$.

\vspace{0.03in}\noindent
{\sl iv}. 1-brane/string BPS systems are truly bound, 
$E = \sqrt{ (N_1E_1)^2 + (N_sE_s)^2}$.
\vspace{0.05in}

The energies given for these states are the exact energies expected
from string theory.  These are expected to correspond with the
Born-Infeld energies of these bound state configurations.  From the
Yang-Mills point of view we only see the $F^2$ term in the expansion
of the Born-Infeld energy
around a flat background, as in (\ref{eq:action-expansion}).
A static field configuration on a single flat $p$-brane has
Born-Infeld energy
\begin{equation}
E_{{\rm BI}} = \tau_p \sqrt{\det (\delta_{ij} + 2 \pi \alpha' F_{ij})}.
\label{eq:Born-Infeld-energy}
\end{equation}
It is not completely understood at this time how to generalize the
Born-Infeld action to arbitrary non-abelian fields
\mcite{Tseytlin,ht}.  In the case where all components of the
field strength commute, however, the Born-Infeld action can be defined by
simply taking a trace outside the square root in
(\ref{eq:Born-Infeld-energy}).  This gives the expected formula for
the non-abelian super Yang-Mills energy at second order
\begin{equation}
E_{{\rm YM}} = \tau_p \pi^2 \alpha'^2 \int {\rm Tr}\; F_{ij}^2.
\label{eq:SYM-energy}
\end{equation}
We will now discuss the descriptions of various bound states in the
super Yang-Mills formalism and show that (\ref{eq:SYM-energy}) indeed
has the expected BPS value for these systems.

{}
\vspace{0.08in}
\noindent
{\it \bzt\ \ \ 0-2 bound states}
\vspace{0.05in}

The simplest bound state of multiple D-branes from the point of view
of Yang-Mills theory is a bound state of 0-branes and 2-branes where
the 2-branes are wrapped around a compact 2-torus \mcite{Townsend}.  As
discussed in Section \ref{sec:examplet2}, a system containing $N$
2-branes and $k$ 0-branes (with the 0-branes confined to the surface
of the 2-branes) is described by a $U(N)$ Yang-Mills theory with total
magnetic flux $\int F = 2 \pi k$.  From simple dimensional
considerations it is clear that the energy of the configuration is
minimized when the flux is distributed as uniformly as possible on the
surface of the 2-branes.  This follows from the fact that in the
Yang-Mills theory the energy scales as $\int F^2$.  For example, if we
consider a field configuration $F$ corresponding to a 0-brane on an
infinite 2-brane, the energy can be scaled by a factor of $\rho^2$
while leaving the flux invariant by taking $F (x) \rightarrow \rho^2 F
(\rho x)$; thus the energy can be taken arbitrarily close to 0 by
taking $\rho \rightarrow 0$.

On a compact space such as $T^2$, the energy is minimized when the
flux is uniformly distributed.  Precisely such a configuration of $N$
2-branes and $k$ 0-branes was
considered in Section \ref{sec:examplet2}.  The Yang-Mills energy of this
configuration corresponds to the second term in the power series
expansion of the expected Born-Infeld energy for a BPS configuration
\begin{equation}
E =\sqrt{(N \tau_2 L_1 L_2)^2 + (k \tau_0)^2}
= N \tau_2 L_1 L_2+ \tau_2 \pi^2 \alpha'^2 \int {\rm Tr}\; F^2 + \cdots
\end{equation}
where 
\begin{equation}
F_{12} = \frac{2 \pi k}{ N L_1L_2}  \identity.
\end{equation}
Thus, we see that the Yang-Mills energy is indeed that expected of a BPS bound
state.  The fact that this configuration is truly bound is
particularly easy to see in the T-dual picture, where it corresponds
to a state of D1-branes with winding numbers $N, k$ on the dual
torus.  Clearly, when $N$ and $k$ are relatively prime, the lowest
energy state of this 1-brane system is a single diagonally wound
brane.  This is precisely the system described in Section \ref{sec:examplet2}
as the dual of the 0-2 system with uniform flux density.  When $N$ and
$k$ have a greatest common denominator $n$ then the system can be
considered to be a marginally bound configuration of $n$ $(N/n, k/n)$
states.  In this case the moduli space of constant curvature solutions
has extra degrees of freedom corresponding to the independent motion of
the component branes \mcite{Sanjaye-Zack2}.

Since the 0-2 bound states saturate the BPS bound on the energy, it is
natural to try to check that there is an unbroken supersymmetry in the
super Yang-Mills theory.  Naively applying the supersymmetry
transformation (\ref{eq:Yang-Mills-SUSY})
\begin{equation}
\delta \psi = -\frac{1}{4}  F_{\mu \nu} \Gamma^{\mu \nu} \epsilon
\end{equation}
it seems that the state is not supersymmetric, since
\begin{equation}
(\Gamma^{12})^2 = -1
\end{equation}
and therefore $\delta \psi$ cannot vanish for all $\epsilon$ when
$F_{12} \sim \identity$.  There is a subtlety here, however
\mcite{Green-Gutperle,Balasubramanian-Leigh}.  In the IIA string theory
there are 32 supersymmetries.  16 are broken by the 2-brane and
therefore do not appear in the SUSY algebra of the gauge theory.  To
see the unbroken supersymmetry it is necessary to include the extra 16
supersymmetries, which appear as linear terms in
(\ref{eq:Yang-Mills-SUSY}).  After including these terms we see that
as long as $F$ is constant and proportional to the identity, the
Yang-Mills configuration preserves 1/2 of the original 32
supersymmetries, as we would expect for a BPS state of this type.
Thus, although the $0-2$ bound state breaks the original 16
supersymmetries of the SYM theory, there exists another linear
combination of 16 SUSY generators under which the state is invariant.

{}
\vspace{0.08in}
\noindent
{\it  \bzf\ \  \ 0-4 bound states}
\vspace{0.05in}

We now consider bound states of 0-branes and 4-branes.  A system of
$N$ 4-branes, no 2-branes and $k$ 0-branes is described by a $U(N)$
Yang-Mills configuration with instanton number $C_2= k$, as discussed in
Section \ref{sec:examplet4}.  Unlike the 0-2 case, on an
infinite 4-brane world-volume the Yang-Mills configuration can be
scaled arbitrarily without changing the energy of the system.  This
follows from the fact that the instanton number and the energy both
scale as $F^2$.  The set of Yang-Mills solutions which minimize the
energy for a fixed value of $C_2$ form the moduli space of $U(N)$
instantons.  This corresponds to the classical moduli space of 0-4
bound states.

If we compactify the 4-brane world-volume on a torus $T^4$ then the
moduli space of 0-4 bound states becomes the moduli space of $U(N)$
instantons on $T^4$ with instanton number $k$ \mcite{Vafa-instantons}.
As an example we now describe a particularly simple class of
instantons in the case $N = k = 2$ considered in Section
\ref{sec:examplet4}.  If we allow the dimensions of the torus to be
arbitrary, there are solutions of the Yang-Mills equations with
constant curvature $F_{12} = 2 \pi \tau_3 \identity/(L_1 L_2), F_{34}
= 2 \pi \tau_3 \identity/(L_3 L_4)$.  It is a simple exercise to check
that the Yang-Mills energy of this configuration is greater or equal
to the energy $2 \tau_0$ of two 0-branes, with equality when $L_1 L_2=
L_3 L_4$.  In fact, in the extremal case the Born-Infeld energy
\begin{equation}
E =  2 \tau_4 V_4  \sqrt{(1 +  4 \pi^2 \alpha'^2 F_{12}^2)
(1 +  4 \pi^2 \alpha'^2 F_{34}^2)} =2 \tau_4   V_4+2 \tau_0
\end{equation}
factorizes exactly so that there are no higher order corrections to
the Yang-Mills energy.  The extremality condition here amounts to the
requirement that the field strength $F$ is self-dual.  In this case,
precisely 1/4 of the supersymmetries of the system are preserved, and
the mass is therefore BPS protected.  As discussed in Section
\ref{sec:examplet4}, this field configuration is T-dual to a
configuration of two 2-branes intersecting at angles.  The
self-duality condition is equivalent to the condition that the angles
$\theta_1, \theta_2$ relating the intersecting branes are equal; this
is precisely the necessary condition for a system of
intersecting branes to preserve some supersymmetry  \mcite{bdl}.

In general, on any manifold the moduli space of instantons is
equivalent to the space of self-dual or anti-self-dual field
configurations.  This follows essentially from the inequality
\begin{equation}
\int(F \pm *F)^2 = 2 \int (F^2 \pm F \wedge F) \geq 0.
\end{equation}
As we have discussed, the moduli space of instantons is, roughly
speaking, the classical moduli space of bound state configurations for
a 0-4 system.  There are several complications to this story, however,
which we now discuss briefly.

The first subtlety is that when an instanton shrinks to a point, the
associated 0-brane can leave the surface of the 4-branes on which it
was embedded.  Although this is a natural process from the string
theory point of view, this phenomenon is not visible in the gauge
theory living on the 4-brane world-volume.  Thus, to address questions
for which this process is relevant, a more general description of a
0-4 system is needed.  One approach which has been used
\mcite{Witten-small,Vafa-instantons}
is to incorporate two
gauge groups $U(N)$ and $U(k)$, describing simultaneously the
world-volume physics of the 4-branes and 0-branes.  In addition to the
gauge fields on the two sets of branes this theory contains a set of
additional hypermultiplets $\chi$ corresponding to 0-4 strings.  If
the dynamics of the 4-brane are dropped by ignoring fluctuations in
the $U(N)$ fields, then the remaining theory is the dimensional
reduction of an ${\cal N} = 2$ super Yang-Mills theory in four
dimensions with $N k$ hypermultiplets.  The moduli space of vacua for
this theory has two branches: a Coulomb branch where $\chi = 0$ and a
Higgs branch where the 0-brane lies in the 4-brane world-volume.  It
was shown by Witten \mcite{Witten-small} (in the analogous context of
5-branes and 9-branes) that the Higgs branch of this theory is precisely
the moduli space of instantons on $\br^4$.  In fact, the ADHM
construction of this moduli space involves precisely the hyperk\"ahler
quotient which gives the Higgs branch of the
moduli space of vacua for the ${\cal N} = 2$ Yang-Mills
theory.  The generalization of this situation to arbitrary $p-(p + 4)$
brane systems was discussed by Douglas \mcite{Douglas,Douglas-gauge}
who also showed that the instanton structure can be seen by a probe
brane.

A second complication which arises in the discussion of 0-4 bound
states is that on compact manifolds such as $T^4$ for certain values
of $N$ and $k$ there are no smooth instantons.  For example, for $N =
2$ and $k = 1$, instantons on $T^4$ tend shrink to a point so there
are no smooth instanton configurations with these quantum numbers.  It
was argued by Douglas and Moore \mcite{Douglas-Moore} that a complete
description of the moduli space in this case requires the more
sophisticated mathematical machinery of sheaves.  Using the language
of sheaves it is possible to describe a moduli space analogous to the
instanton moduli space for arbitrary $N, k$.  One argument for why
this language is essential is that the Nahm-Mukai transform which
gives an equivalence between moduli spaces of instantons on the torus
with $(N, k)$ and $(k, N)$ is only defined for arbitrary $N$ and $k$
in the sheaf context (See \mycite{Donaldson-Kronheimer} for a review
of the Nahm-Mukai transform and further references).  This equivalence
amounts to the statement that the moduli space of 0-4 bound states is
invariant under T-duality, which is a result clearly expected from
string theory.

This discussion has centered around the classical moduli space of 0-4
bound states.  In the quantum theory, the construction of bound states
essentially involves solving supersymmetric quantum mechanics on this
moduli space, giving a relationship between the number of discrete
bound states and the cohomology of the moduli space \mcite{Vafa-gas}.
Precisely solving this counting problem requires understanding how the
singularities in the moduli space are resolved quantum mechanically.
The mathematics underlying the resolution of these singularities again
involves sheaf theory \mcite{Harvey-Moore,Nakajima}.  A fully detailed
description of how this state counting problem works out on a general
compact 4-manifold has not been given yet, although
there are many results in special cases, particularly for asymptotic
values of the charges, which are applicable to entropy analysis for
stringy black holes; this issue will be discussed in further detail in
the lectures of Maldacena at this school.

{}
\vspace{0.08in}
\noindent
{\it \bzs\ \  \ 0-6 and 0-8 bound states}
\vspace{0.05in}

So far we have discussed 0-2 and 0-4 bound states from the Yang-Mills
point of view.  In both cases there are classically stable Yang-Mills
solutions which correspond to a $p$-brane with a gauge field strength
carrying 0 brane charge.  It is natural to ask what happens when we
try to construct analogous configurations for $p = 6$ or 8.  From the
scaling argument used above, it is clear that a 0-brane on an infinite
6- or 8-brane will tend to shrink to a point, since the 0-brane charge
scales as $F^3$ or $F^4$ while the energy scales as $F^2$.  Thus, in
general we would expect that a 0-brane spread out on the surface of a
6- or 8-brane would tend to contract to a point and then leave the
surface of the higher dimensional brane.  In fact, analysis of the
SUSY algebra in string theory indicates that BPS states containing
0-brane and 6- or 8-brane charge have vanishing Yang-Mills energy so
that the 0-brane cannot have nonzero size on the 6/8-brane.
Strangely, however,  on the torus $T^6$ or $T^8$ there
are (quadratically) stable Yang-Mills configurations with charges
corresponding to 0-branes and no other lower-dimensional branes
\mcite{WT-adhere}.  For example, on $T^6$ we can construct a field
configuration with
\begin{equation}
F_{12}  =  2 \pi \mu_1\;\;\;\;\;
F_{34}  =  2 \pi \mu_2\;\;\;\;\;
F_{56}  = 2 \pi \mu_3
\end{equation}
where
\begin{equation}
\mu_1=\left(
\begin{array}{cccc}
 1 & 0 & 0 & 0\\
0 & 1 & 0 & 0\\
0 & 0 & -1 & 0\\
0 & 0 & 0 & -1\\
\end{array} \right)
\;\;\;\;\; 
\mu_2=\left(
\begin{array}{cccc}
 1 & 0 & 0 & 0\\
0 & -1 & 0 & 0\\
0 & 0 & -1 & 0\\
0 & 0 & 0 & 1\\
\end{array} \right)
\end{equation}
\begin{equation}
\mu_3=\left(
\begin{array}{cccc}
 1 & 0 & 0 & 0\\
0 & -1 & 0 & 0\\
0 & 0 & 1 & 0\\
0 & 0 & 0 & -1\\
\end{array} \right)
\end{equation}
This solution is quadratically stable, but breaks all supersymmetry.
It is T-dual to a system of 4 3-branes intersecting pairwise on
lines.  In the quantum theory these configurations must be unstable
and will eventually decay; however, because of the classical quadratic
stability we might expect that the states would be fairly long-lived.
These configurations seem to be related to metastable
non-supersymmetric black holes \mcite{Khuri-Ortin,Sheinblatt}.

{}
\vspace{0.08in}
\noindent
{\it \ppb\ \ \  $p-p$ bound states}
\vspace{0.05in}


We now consider the question of bound states between parallel branes
of the same dimension.  As in the case of 0-4 bound states, the
existence of $p-p$ bound states depends crucially upon subtleties in
the quantum theory, and is a somewhat complicated question.  We review
the story here very briefly; for a much more detailed analysis the
reader is referred to the paper of Sethi and Stern \mcite{Sethi-Stern}.
Recall that the world-volume theory of $N$ $p$-branes is $U(N)$ ${\cal
N} = 1$ SYM in 10D dimensionally reduced to $p + 1$ dimensions.  The
bosonic fields in this theory are $A_{\alpha}$ and $X^a$.  The moduli
space of classical vacua for $N$ $p$-branes is the configuration space
\begin{equation}
\frac{(\br^{9-p})^N}{S_N} 
\end{equation}
where $S_N$ arises from Weyl symmetry.  Thus, classically the branes
move freely and there is no apparent reason for a bound state to
occur.

Once we include quantum effects, the story becomes more subtle.  Let
us restrict ourselves for simplicity to the case of two 0-branes,
corresponding to supersymmetric $U(2)$ matrix quantum mechanics
(\ref{eq:super-qm}).  If we had a purely bosonic theory, then it is
easy to see that if we consider a classical configuration where the
two 0-branes are separated by a distance $r$ then the off-diagonal
matrix elements would behave like harmonic oscillators with a quantum
ground state energy proportional to $r$.  Thus, in the classical
bosonic theory we expect to see a discrete spectrum of bound
states \mcite{Simon}.  Once supersymmetry has been included, the fermions
contribute ground state energies with the opposite sign, which
precisely cancel the zero-point energies of the bosons.  In principle,
this allows for the possibility of a zero-energy ground state
corresponding to a marginal bound state of two 0-branes.  The
existence of such a state was finally proven definitively in the work
of Sethi and Stern \mcite{Sethi-Stern}.  Remarkably, the existence of
the marginally bound state depends crucially upon the large degree of
supersymmetry in the 0-brane matrix quantum mechanics.  The analogous
theories with 8 and 4 supersymmetries which arise from dimensional
reduction of ${\cal N} = 1$ theories in 6 and 4 dimensions have no
such bound states.

{}
\vspace{0.08in}
\noindent
{\it \bsdf\ \ \ Bound states of D-strings and fundamental strings}
\vspace{0.05in}

We conclude our discussion of bound states with a brief discussion of
bound states of D-strings with fundamental strings.  This was in fact
the first of the bound state configurations described here to be
analyzed from the point of view of the D-brane world-volume gauge
theory \mcite{Witten-bound}.  In IIB string theory, states of
1-branes and strings with quantum numbers $(N, q)$ transform as a
vector under the $SL(2,Z)$  S-duality symmetry.  Combining this
symmetry with T-duality, the following diagram shows that $N$
D-strings and $q$ fundamental strings should form a truly bound state
when $N$ and $q$ are relatively prime since this configuration is dual
to  an $N, q$ 2-0 system:
\begin{center}
\centering
\centering
\begin{picture}(200,40)(- 100,- 25)
\put(-120,0){\makebox(0,0){IIA}}
\put(120,0){\makebox(0,0){IIB}}
\put(-40,0){\makebox(0,0){IIB}}
\put(40,0){\makebox(0,0){IIB}}
\multiput(-108,0)(80,0){3}{\vector(1,0){56}}
\multiput(-52,0)(80,0){3}{\vector(-1,0){56}}
\put(-80,10){\makebox(0,0){\tiny $T_3$}}
\put(0,10){\makebox(0,0){\tiny $S$}}
\put(80,10){\makebox(0,0){\tiny $T_{12}$}}
\put(-120,-15){\makebox(0,0){\small D2(12)+D0}}
\put(-40,-15){\makebox(0,0){\small D3(123)+D1(3)}}
\put(40,-15){\makebox(0,0){\small D3(123)+F1(3)}}
\put(120,-15){\makebox(0,0){\small D1(3)+F1(3)}}
\end{picture}
\end{center}
Indeed, Witten showed that an argument for the existence of such a
bound state can be given in terms of the world-volume
gauge theory.  As discussed in Section
\ref{sec:strings-electric}, string winding number is proportional to
electric flux on the 1-brane world-sheet.  As mentioned previously,
the quantization of fundamental string number is therefore a quantum
effect in this gauge theory; the flux quantum in the $U(N)$ theory
\begin{equation}
e = \frac{g}{2 \pi \alpha' N} 
\end{equation}
can be related to the fundamental unit of momentum of a 0-brane on a
T-dual circle
\begin{equation}
\pi = \frac{\dot{x}}{ \hat{g} \sqrt{\alpha'}}  = \frac{1}{N\hat{R}} 
\end{equation}
through
\begin{equation}
\hat{g} \sqrt{\alpha'} \pi = (2 \pi \alpha') e = g/N.
\end{equation}
In terms of this flux quantum it
is easy to check that the leading term in the Born-Infeld expression
for the bound state energy indeed corresponds with that found in the
gauge theory 
\begin{eqnarray}
E & 
= & L \sqrt{(\tau_1 N)^2 + (\frac{k}{2 \pi \alpha'})^2 } = L \tau_1 N
\sqrt{1 + k^2 g^2/N^2} \nonumber\\
& = & L \tau_1 N +\frac{1}{2} L \tau_1
(4 \pi^2 \alpha'^2 k^2 e^2) + \cdots
\end{eqnarray}
{}
\normalsize

\subsection{Potentials between separated D-branes}

\begin{figure}
\vspace{-0.3in}
\psfig{figure=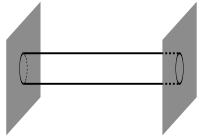,height=1.5in}
\vspace{-0.3in}
\caption[x]{\footnotesize Annulus diagram for D-brane interactions}
\label{f:annulus}
\end{figure}

Now that we have discussed bound states of various types of D-branes,
we go on to consider interactions between separated branes.  In string
theory the dominant long-distance interaction between D-branes is
found by calculating the annulus diagram which
corresponds to the exchange of a closed string between the two objects
(See Figure~\ref{f:annulus}).  At long distances, this amplitude is
dominated by the massless closed string modes, which give an effective
supergravity interaction between the objects.  The annulus diagram can
also be interpreted in the open string channel as a one-loop diagram.
We expect that the gauge theory description of the interaction between
the branes should be given by the massless open string modes, which
are relevant at short distances.  These two calculations (restricting
to massless closed and open strings respectively) represent different
truncations of the full string spectrum.  There does not seem to be
any {\it a priori} reason why gauge theory should correctly describe long
distance physics, although in some cases the calculations may agree
because the configuration has some supersymmetry protecting its
physical properties.  The original computations of interactions
between separated D-branes were carried out in the context of the full
string theory spectrum \mcite{Polchinski,Bachas,Lifschytz}.  In
the spirit of these lectures,  however, we will confine our discussion to
aspects of D-brane interactions which can be studied in the context of
Yang-Mills theory.  As we shall see, many of the important qualitative
features (and some quantitative features) of D-brane interactions can
be seen from this point of view.

In the next few subsections we consider static potentials between
branes of various dimensions.  Using T-duality, one of these branes
can always be transformed into a 0-brane, so without loss of
generality we restrict ourselves to interactions between 0-branes and
$p$-branes with $p$ even.

{}
\vspace{0.08in}
\noindent
{\it \spp\ \  \ Static $p-p$ potential}
\vspace{0.05in}

To begin with, let us consider a pair of parallel $p$-branes.  In
Yang-Mills theory such a configuration is described by a $U(2)$ gauge
theory with a nonzero scalar VEV
\begin{equation}
\langle X^a\rangle = d (\tau_3 +  \identity)/2 = \left(\begin{array}{cc}
d & 0\\
0 & 0
\end{array} \right)
\end{equation}
where $d$ is the distance between the branes.  For any $d$, this is a
BPS configuration with Born-Infeld energy 2 $E_p$ and vanishing
Yang-Mills energy.  Therefore there is no force between the branes
even in the quantum theory.  This agrees with the results of the full
string calculation by Polchinski \mcite{Polchinski}.  In the (closed)
string
calculation, there is a delicate balance between NS-NS and R-R string
exchanges.  Note that in a purely bosonic theory, although there is no
classical potential between the branes, there is a quantum-mechanical
attraction between the branes due to the zero-point energy of the
off-diagonal fields, as mentioned in the discussion of 0-brane bound
states.

{}
\vspace{0.08in}
\noindent
{\it\szt\ \  \ Static 0-2 potential \footnote{This subsection is based
on conversations with Hashimoto, Lee, Peet and Thorlacius.}}
\vspace{0.05in}

\begin{figure}
\vspace{-0.3in}
\psfig{figure=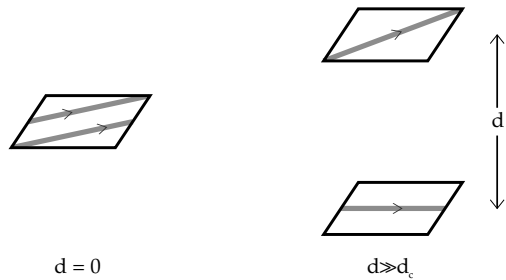,height=2in}
\vspace{-0.3in}
\caption[x]{\footnotesize  D-string configurations which are T-dual to
a pair of separated 2-branes with a unit of flux (0-brane charge)}
\label{f:0-2}
\end{figure}

Unlike the $p-p$ system, a configuration containing a single 2-brane
and a single 0-brane cannot be described by a simple gauge theory
configuration when the branes are not coincident.  This makes it
slightly more difficult to study the interactions between a 0-brane
and a 2-brane in Yang-Mills theory.  Since we know that the
static potential between a pair of 2-branes must vanish, however, we can study
the static potential between a 0-brane and a 2-brane by attaching the
0-brane to an auxiliary 2-brane.  Thus, we consider a pair of 2-branes
on a torus $T^2$ of area $L^2$, corresponding to a $U(2)$ gauge theory
with a single unit of magnetic flux ${\rm Tr}\int F = 2 \pi$.  If we
fix the expectation values of the scalar fields $X^a$ to vanish except
in a single direction with $X^3 = d \tau_3/2$ then the branes are
fixed at a relative separation $d$.  For a fixed value of $d$, we can
then minimize the Yang-Mills energy associated with the gauge field
$A_\alpha$.  This energy will depend upon the separation $d$ because of the
terms of the form $[A_\alpha, X^3]^2$, and gives a classical potential
function $V (d)$.  As we discussed in Section \bzt, when $d = 0$ the
energy will be minimized when the flux is shared between the two
2-branes, corresponding to a (2,1) bound state.  In this case the
Yang-Mills energy of the system is proportional to
\begin{equation}
v (0) = \frac{L^2}{4 \pi^2} \int  {\rm Tr}\; F_{(0)}^2 = 1/2
\end{equation}
On the other hand, when $d$ is very large the energy
will be minimized when the flux is constrained to one of the diagonal
U(1)'s corresponding to a single brane, so that
\begin{equation}
v (\infty) =\frac{L^2}{4 \pi^2}\int  {\rm Tr}\; F_{(\infty)}^2 = 1.
\end{equation}
In this case the flux cannot be shared because the constant scalar
field $X^3$ is not compatible with the boundary conditions
(\ref{eq:boundary2}) needed for the curvature to be proportional to
the identity matrix.  The energetics of these two limits are easy to
understand in the T-dual picture, where the configuration at $d = 0$
corresponds to a single diagonally wrapped (2,1) D-string while the $d
\rightarrow \infty$ configuration corresponds to a pair of strings
with windings (1, 0) and (1,1) (See Figure~\ref{f:0-2}).

In fact, it turns out that the potential function $v (d)$ is constant
for any $d$ greater than a critical distance $d_c$ \mcite{vb-unstable}.
Beyond this distance, the 0-brane and 2-brane have essentially no
interaction classically.  Below this distance, however, the solution
where the flux is confined to a single 2-brane becomes unstable and
the potential function drops continuously down to the value 1/2 at $d
= 0$.  When quantum effects are included, for example by integrating
out at one loop the off-diagonal terms, the potential is smoothed and
the force between the objects becomes nonzero and attractive at
arbitrary distance.  These results agree perfectly with the full
string calculation, which indicates that there is an
attractive potential at all distances \mcite{Lifschytz}.

It is interesting to compare this analysis with a similar discussion
by Douglas, Kabat, Pouliot and Shenker  (DKPS) \mcite{DKPS} of the
potential between a pair of 1-branes carrying a single fundamental
string charge.  The situation they consider is dual to the 0-2
configuration we are discussing; however, because the quantization of
electric field strength occurs only at the quantum level, the
potential they calculate appears only when quantum effects are
considered.  The one-loop potential they calculate is smooth and gives
a nonzero attractive force at all distances, as we expect from the
one-loop calculation in the 0-2 case.

The fact that the force between the 0-brane and 2-brane is mostly
localized within a finite distance $d_c$ provides a simple example of
a general feature which is most clearly seen in brane-anti-brane
interactions \mcite{Banks-Susskind}.  Namely, when two brane
configurations are separately stable but can combine to form a lower
energy configuration, at a distance analogous to $d_c$ a tachyonic
instability appears in the system which indicates the existence of the
lower energy configuration.  A similar situation to the one we have
described here occurs when two 2-branes are provided with 0-brane and
anti-0-brane charges respectively.  In this case when the 2-branes are
brought sufficiently close a tachyonic instability appears which
allows the  0-brane and anti-0-brane  to annihilate.  In the
Yang-Mills language we are using here, these unstable modes can be
explicitly constructed as degrees of freedom in the gauge field.
Because of the nontrivial boundary conditions on the field,
these degrees of freedom are described as theta functions which are
sections of a nontrivial U(1) bundle on the torus.  Using these theta
functions, the tachyonic instabilities associated with brane-brane and
brane-anti-brane forces can be precisely analyzed \mcite{ht,gns}.

{}
\vspace{0.08in}
\noindent
{\it \szf\ \  \ Static  0-4 potential}
\vspace{0.05in}

Just as we placed a 0-brane on an auxiliary 2-brane to determine the
form of the static 0-2 potential, we can place a 0-brane on a pair of
auxiliary 4-branes to determine the static potential between a 0-brane
and a 4-brane.  Thus, we consider a set of 3 (uncompactified) 4-branes
with a scalar field $X^5 = {\rm Diag} (d, d, 0)$, with a $U(2)$
instanton living on the first two 4-branes.  This configuration has a
self-dual gauge field, so it is a BPS state in the moduli space of
$U(3)$ instantons.  Thus, the potential is independent of the distance
$d$ even after quantum effects are considered and we see that there is
no static potential between 0-branes and 4-branes.  This is in
agreement with the results from the full string calculation
\mcite{Lifschytz}.

\vspace{0.08in}
\noindent
{\it \szs\ \  \ Static 0-6  and 0-8 potential}
\vspace{0.05in}

The static potential between a 0-brane and a 6-brane or an 8-brane is
not as easy to understand from the point of view of gauge theory as in
the 0-0, 0-2, and 0-4 cases, since there are no known 0-6 or 0-8 bound
states.  As mentioned in Section \bzs, however, a set of 4 or 8
0-branes can be smoothly distributed on the world-volume of 4 or 8
6-branes or 8-branes in a stable way after energy is added to the
system.  In the case of the 6-brane, this corresponds to the fact that
there is a repulsive interaction between 0-branes and 6-branes (as
determined by the string calculation), so that energy is needed to
push them together from an infinite separation.  Based on Yang-Mills
theory alone, then, one might think that in the 0-8 case one would
also get a repulsive force between the branes.  There is an extra
complication in this case, however, arising from interactions via R-R
fields.  In fact, the potential between separated 0-branes and
8-branes vanishes and such configurations preserve some supersymmetry.
The 0-8 story has a number of subtleties, however.  For example, as a
0-brane passes through an 8-brane, a (fundamental) string is created
\mcite{Hanany-Witten,bdg,dfk,bgl}.  This string produces a charge
density on the 8-brane world-volume.  The physics associated with this
system is still a matter of some discussion.

\vspace{0.08in}
\noindent
{\it \sczz\ \ \ 0-brane scattering}
\vspace{0.05in}

We will now consider the interaction between a pair of moving
0-branes.  The classical configuration space for a pair of 0-branes is
the flat quotient space \begin{equation} \frac{(\br^9)^2}{{\bz}_2}.
\end{equation} As discussed in Section \spp, this configuration space is
protected by supersymmetry, so that all points in the space correspond
to classical BPS states of the two 0-brane system.  When the two
0-branes have a nonzero relative velocity, however, the supersymmetry
of the system is broken and a velocity-dependent potential appears
between the branes.  The leading term in this potential can be
determined by performing a one-loop calculation in the 0-brane quantum
mechanics theory.  We will now review this calculation briefly.  The
Yang-Mills calculation of the potential between two moving 0-branes
was first carried out by Douglas, Kabat, Pouliot and Shenker
\mcite{DKPS}; many variations on this calculation have appeared in the
literature over the last year or so, particularly in the context of
Matrix theory.  This calculation will be discussed further in
Section  \ref{sec:matrix-interactions}.

To find the velocity-dependent potential at one-loop, we begin by
considering a classical background solution for the two-particle
system in which the two 0-branes are moving with relative velocity $v$
in the $X^1$ direction with an impact parameter of $b$ along the $X^2$
axis
\begin{eqnarray}
X^2 (t) & = &\left(\begin{array}{cc}
b & 0\\
0 & 0
\end{array} \right)\\
X^1 (t)  &= & \left(\begin{array}{cc}
vt & 0\\
0 & 0
\end{array} \right)  \nonumber
\end{eqnarray}

To calculate the effective potential between these 0-branes at
one-loop order, we need to integrate out the off-diagonal fields.  We
can perform the calculation in background-field gauge, where we set
\begin{equation}
X^a = \langle X^a \rangle + \delta X^a
\end{equation}
and where we add a background-field gauge fixing term
\begin{equation}
-\frac{1}{2 R}  \left(\dot{A}_0+i[\langle X^a \rangle, X^a] \right)^2
\end{equation}
to the Lagrangian (\ref{eq:super-qm}).  We can calculate the one-loop
potential by expanding the action to quadratic order in the
off-diagonal fluctuations.  In the quasi-static approximation, which
is valid to leading order in the inverse separation
\mcite{Tafjord-Periwal}, the potential is then simply given by the sum
of the ground state energies of the corresponding harmonic
oscillators.  There are 10 (complex) bosonic oscillators, with frequencies
\begin{eqnarray*}
\omega_b & = &  \sqrt{r^2} \;\;\;\;\; {\rm with\ multiplicity\ 8}\\
\omega_b & = &  \sqrt{r^2 \pm 2iv} \;\;\;\;\;
{\rm with\ multiplicity\ 1\ each}.
\end{eqnarray*}
where $r = \sqrt{b^2 + v^2 t^2}$ is the distance between the branes at
time $t$.
There are also 2 ghosts with frequencies 
\begin{equation}
\omega_g = \sqrt{r^2},
\end{equation}
and there are 16 fermions with frequencies
\begin{equation}
\omega_f = \sqrt{r^2 \pm iv} \;\;\;\;\;
{\rm with\ multiplicity\ 8\ each}.
\end{equation}
The velocity-dependent potential is then given by
\begin{equation}
V = \sum_{b}\omega_b-2 \omega_g-1/2 \sum_{f} \omega_f.
\label{eq:oscillator-sum}
\end{equation}
For $v = 0$ the frequencies clearly cancel and the potential
vanishes.  For nonzero $v$ we can expand each frequency in a power
series in $1/r$.  At the first three orders in $v/r^2$ the potential
vanishes; the first nonvanishing term appears at fourth order, so that
the potential between the 0-branes is given at leading order by
\begin{equation}
V (r) =\frac{-15v^4}{16\;r^7}.
\label{eq:scattering-potential}
\end{equation}
As we will discuss in more detail in the following sections, it can be
checked that this is in precise agreement with the corresponding
potential in supergravity, including the multiplicative constant.


{}
\normalsize

\section{M(atrix) theory: The Conjecture}
\label{sec:conjecture}

In the first four lectures we accumulated a fairly wide range of
results which can be derived from the Yang-Mills description of
D-branes.  The last lecture (Sections \ref{sec:conjecture},
\ref{sec:evidence} and \ref{sec:developments}) contains an
introduction to the Matrix theory conjecture, which states that the
large $N$ limit of the Yang-Mills quantum mechanics of 0-branes
contains a large portion of the physics of M-theory and string theory.
As we shall see, much of the evidence for the Matrix theory
conjecture is based on properties of the Yang-Mills description of
D-branes which we have discussed in the context of type II string
theory.  The discussion given here of Matrix theory is fairly
abbreviated and focuses on understanding how the objects and
interactions of supergravity can be found in Matrix theory.  The
core of the material is based on the original lectures given in June
of 1997; however, some more recent material is included which is
particularly germane to the subject matter of the original lectures.
Many important and interesting aspects of the theory are mentioned
briefly, if at all.  Other reviews which discuss some recent
developments in more detail have been given by Banks
\mcite{banks-review} and by Susskind \mcite{Susskind-review}.

This section contains the statement of the Matrix theory conjecture
as well as a brief review of some background material useful in
understanding the statement of the conjecture, namely short reviews of
M-theory and the infinite momentum frame.  In Section
\ref{sec:evidence} we discuss some of the evidence for Matrix
theory, and in Section \ref{sec:developments} we discuss some further
directions in which this theory has been explored.

\subsection{M-theory}

The concept of M-theory has played a fairly central role in the
development of the web of duality symmetries which relate the five
string theories to each other and to supergravity
\mcite{Hull-Townsend,Witten-various,dlm,Schwarz-m,Horava-Witten}.
M-theory is a conjectured eleven-dimensional theory whose low-energy
limit corresponds to 11D supergravity \mcite{cjs}.  Although there are
difficulties with constructing a quantum version of 11D supergravity,
it is a well-defined classical theory with the following field content:
\vspace{0.03in}

\noindent
$e^a_I$: a vielbein field (bosonic, with 44 components)\\
\noindent
$\psi_I$: a Majorana fermion gravitino (fermionic, with 128
components)\\
\noindent
$A_{I J K}$: a 3-form potential (bosonic, with 84 components).
\vspace{0.02in}

In addition to being a local theory of gravity with an extra 3-form
potential field, M-theory also contains extended objects.  These
consist of a two-dimensional supermembrane and a 5-brane.

One way of defining M-theory is as the strong coupling limit of the
type IIA string.  The IIA string theory is taken to be equivalent to
M-theory compactified on a circle $S^1$, where the radius of
compactification $R$ of the circle in direction 11 is related to the
string coupling $g$ through $R= g^{2/3}l_p = g l_s$, where $l_p$ and
$l_s = \sqrt{\alpha'}$ are the M-theory Planck length and the string
length respectively.  The
decompactification limit $R \rightarrow \infty$ corresponds then to
the strong coupling limit of the IIA string theory.  (Note that we
will always take the eleven dimensions of M-theory to be labeled $0,
1, \ldots, 8, 9, 11$; capitalized roman indices $I, J, \ldots$ denote
11-dimensional indices).

Given this relationship between compactified M-theory and  IIA
string theory, a correspondence can be constructed between various objects in
the two theories.  For example, the Kaluza-Klein photon associated
with the components $g_{\mu 11}$ of the 11D metric tensor can be
associated with the R-R gauge field $A_\mu$ in IIA string theory.  The
only object which is charged under this R-R gauge field in IIA string
theory is the 0-brane; thus, the 0-brane can be associated with a
supergraviton with
momentum $p_{11}$ in the compactified direction.  The membrane and
5-brane of M-theory can be associated with different IIA objects
depending on whether or not they are wrapped around the compactified
direction;  the correspondence between
various M-theory and IIA objects is given in Table~\ref{tab:m2}.
\begin{table}[t]
\caption{Correspondence between objects in M-theory and
IIA string theory\label{tab:m2}}\vspace{0.4cm}
\begin{center}
\begin{tabular}{|l|l|}
\hline
M-theory & IIA\\
\hline
KK photon ($g_{\mu 11}$) & RR gauge field $A_\mu$\\
supergraviton with $p_{11}= 1/R$ & 0-brane\\
wrapped membrane & IIA string\\
unwrapped membrane & IIA D 2-brane\\
wrapped 5-brane & IIA D 4-brane\\
unwrapped 5-brane & IIA NS 5-brane\\
\hline
\end{tabular}
\end{center}
\end{table}

\subsection{Infinite momentum frame}

Roughly speaking,
the infinite momentum frame (IMF) is a frame in which the physics  has
been
heavily boosted in one particular direction.  This frame has the
advantage that it simplifies many aspects of relativistic quantum
field theories \mcite{Kogut-Susskind}.
To study a theory in the IMF, we begin by choosing a longitudinal
direction; this will be $X^{11}$ in the case of M-theory.  We then
restrict attention to states which have very large values of momentum
$p_{11}$ in the longitudinal direction.  This is sometimes stated in
the form that any system of interest should be heavily boosted in the
longitudinal direction; however, this latter formulation leads to some
subtleties, particularly when the longitudinal direction is compact.
The basic idea of the IMF frame is that if we are interested in
scattering amplitudes where the in-states and out-states have large
values of $p_{11}$ then we can integrate out all the states with
negative or vanishing $p_{11}$, giving a simplified theory.  In
general, intermediate states without large $p_{11}$ will indeed be
highly suppressed.   Degrees of freedom associated with
zero-modes  can cause complications, however \mcite{Hellerman-Polchinski}.

One advantage of the IMF frame is that it turns a relativistic
theory into one with a simpler, Galilean, invariance group.  If a
state has a large longitudinal momentum $p_{11}$ then to leading order
in $1/p_{11}$ a Lorentz boost acts as a 
Galilean boost on the transverse momentum $p_\bot$ of the state
\begin{equation}
p_\bot \rightarrow p_\bot + p_{11} v_\bot.
\end{equation}
A massless particle has an energy which is given to leading order in
$1/p_{11}$ by 
a Galilean energy
\begin{equation}
E-p_{11} = \frac{p_\bot^2}{2p_{11}} 
\end{equation}
in the IMF.  Thus, we see that the longitudinal momentum $p_{11}$
plays the role of the mass in the IMF Galilean theory.

If the longitudinal direction $X^{11}$ is compact with radius $R$,
then longitudinal momentum is naturally quantized in units of $1/R$,
so that $p_{11} = N/R$.  Note that, as mentioned above, there are
subtleties with boosting a compactified theory; in particular, a boost
is not a symmetry of a Lorentz invariant theory which has been
compactified in the direction of the boost, since after the boost the
constant time surface becomes noncompact.  By simply treating the IMF
frame as a way of calculating interactions between states with large
longitudinal momentum, however, this complication should not concern
us particularly.

The description of a theory in the
infinite momentum frame is closely related to the description of the
theory given  in
light-front coordinates.
In fact, for comparing Matrix theory to supergravity it is most
convenient to use the language of discrete light-front quantization
(DLCQ) \mcite{Susskind-DLCQ}.  In this framework a system is
compactified in a lightlike dimension $x^-$ so that longitudinal
momentum $p^+$ is quantized in units $N/R$, where we set
\begin{equation}
x^\pm = \frac{1}{ \sqrt{2}}  (x^0 \pm x^{11}), \;\;\;\;\;
x^-\equiv x^- + 2 \pi R
\label{eq:}
\end{equation}
(note that the light-front metric has $\eta_{+ -}= -1, \eta_{+ +} =
\eta_{--}= 0$).  The DLCQ prescription gives a light-front description
of a theory in the IMF when $N \rightarrow \infty$.  Further
discussion of DLCQ quantization in the context of Matrix theory can be
found in
\mycite{Sen,Seiberg-DLCQ,Hellerman-Polchinski,Bigatti-Susskind,Susskind-review}.

\subsection{The conjecture}

The following conjecture was made by Banks, Fischler, Shenker and
Susskind (BFSS) \mcite{BFSS}:

{\it M-theory in the IMF is exactly described by the $N \rightarrow
\infty$ limit of 0-brane quantum mechanics
\begin{equation}
{\cal L}= \frac{1}{2 R}  {\rm Tr}\; \left[
  \dot{X}^a \dot{X}_a
+ \sum_{ a < b}[X^a, X^b]^2+  \theta^T (i\dot{\theta}
- \Gamma_a[X^a, \theta]) \right]
\label{eq:matrix-Lagrangian}
\end{equation}
where $N/R$ plays the role
of the longitudinal momentum, and where $N/R$ and $R$ are both taken to
$\infty$.}
Note that (\ref{eq:matrix-Lagrangian}) is the same as
(\ref{eq:super-qm}) in units where $2 \pi \alpha' = 1$, after
replacing $g \sqrt{\alpha'}= R$

Although we will continue to work in the string units in which
(\ref{eq:matrix-Lagrangian}) is expressed, in many references the
Lagrangian is expressed in Planck units
\begin{equation}
{\cal L}={\rm Tr}\;  \left[
\frac{1}{2 R}  \dot{X}^a \dot{X}_a
+ \frac{R}{8\pi^2} \sum_{a < b} [X^a, X^b]^2+ \frac{1}{4 \pi} \theta^T
(i\dot{\theta} 
- \frac{R}{2\pi}  \; \Gamma_a[X^a, \theta]) \right].
\label{eq:matrix-Lagrangian-Planck}
\end{equation}
The change of units can be carried out
by simply replacing $\alpha' \rightarrow l_p^2 g^{-2/3}$ in
(\ref{eq:super-qm}) and setting $l_p = 1$.

The original evidence for this conjecture included the following:

\noindent
$\circ$ Only 0-branes carry $p_{11}$.  Not only does this mean that
states in M-theory with large $p_{11}$ are composed primarily of
0-branes, but this also fits naturally into the holographic principle
espoused by 't Hooft and Susskind
\mcite{Hooft-reduction,Susskind-holographic} which states that at large
momentum string theory states can be described in terms of elementary
partons which each take up a single Planck unit of transverse area.
(Related ideas have also been discussed by Thorn \mcite{Thorn-bits}.)

\noindent
$\circ$ The 10D Super-Galilean invariance of (\ref{eq:matrix-Lagrangian}).

\noindent
$\circ$ The fact that
graviton scattering amplitudes in 11D supergravity are correctly
described by the scattering amplitude of 0-branes arising from the leading
$v^4/r^7$ potential term.

\noindent
$\circ$ The natural appearance of the supermembrane in the matrix
quantum mechanics theory \mcite{dhn}.  This connection between the
low-energy theory of 0-branes and the light-front supermembrane theory
was also pointed
out by Townsend \mcite{Townsend}.

In the time since this conjecture was made, supporting evidence
has continued to appear.  In the following section, we will discuss
some of this evidence.  

{}

\subsection{Matrix compactification}
\label{sec:compactification}

Before discussing in detail the evidence for Matrix theory, let us
discuss briefly the issue of compactifying the theory.  Compactifying
Matrix theory on a manifold $M$ would correspond to a
compactification of M-theory on $M \times S^1$ where the $S^1$ is
taken to a decompactified limit through $R \rightarrow \infty$.  There
are several ways in which BFSS suggested it might be possible to
define a compactified version of Matrix theory.

The first approach to compactifying the theory would be to simply
define Matrix theory on a manifold $M$ to be the large $N$ limit of
the theory of $N$ 0-branes on $M$.  For example, by using the
equivalence discussed in Section \ref{sec:T-duality} between the 0-brane
theory on a torus and super Yang-Mills theory on the dual torus, this
would define Matrix theory on the torus $T^d$ in terms of a
$d$-dimensional super Yang-Mills theory.  The torus can then be modded
out by a finite group to get Matrix theory on an orbifold.
So far 0-brane quantum mechanics is only very well understood
on tori and orbifolds, however.  There has been some progress made on curved
manifolds, particularly on K3 and Calabi-Yau spaces
\mcite{dos,Douglas-Ooguri}, however 
the situation is not as clear in these cases.  Thus, this approach
does not immediately lead to a candidate definition of Matrix theory
compactified on an arbitrary manifold.

A second approach to compactifying Matrix theory 
involves taking superselection sectors of the theory which may
correspond to different compactifications.  For example, in the large
$N$ limit we can take infinite matrices satisfying
\begin{equation}
U X^a U^{-1} = X^a + \delta^{a9} 2 \pi R_9\identity.
\end{equation}
for some ``translation'' operator $U$ and radius $R_9$.  This
superselection sector of the theory corresponds to an $S^1$
compactification, since the matrices satisfying this relation can be
interpreted in terms of the fields of $(1 + 1)$-D super Yang-Mills theory on
the circle as in (\ref{eq:constraint2}).  In a similar way, it is easy
to see that Matrix theory ``contains'' the SYM theory in all
dimensions $d \leq 10$.  It is an interesting open question whether
there are other superselection sectors of the theory which naturally
correspond to compactifications on non-toroidal spaces.

Both of these approaches to Matrix theory compactification give the
same prescription for compactifying the theory on a torus.  We will
use this description of Matrix theory on a torus in terms of the super
Yang-Mills theory on the dual torus in the following discussion.  For
compactification on tori of dimension $d > 3$, however, additional
features emerge which make the story more complicated.  This issue will be
discussed briefly in Section \ref{sec:compactification2}.

\section{Matrix theory: Symmetries, Objects and Interactions}
\label{sec:evidence}

If the Matrix theory conjecture is correct, we would expect that all
the symmetries of M-theory should correspond  to symmetries of
(\ref{eq:matrix-Lagrangian}).  Furthermore, it should be possible to
find matrix constructions of all the objects we expect to see in the
11-dimensional  supergravity theory which describes M-theory at low
energies, and the interactions between these objects in Matrix theory
should agree with the interactions between the corresponding
supergravity objects.
Most of the evidence to date for the Matrix theory
conjecture consists of showing that some piece of this correspondence
is correct.  In this section we review some of this evidence, divided
into the three categories mentioned.  Recent arguments for the Matrix
theory conjecture based on more general principles are discussed in
Section \ref{sec:recent}.

\subsection{Symmetries in Matrix theory}

There are two important symmetries of M-theory which we would like to
see reproduced in Matrix theory.  First is the Lorentz symmetry of the
theory.  This is explicitly broken in the IMF; nonetheless, one would
hope that a residual version of this symmetry would still be present.
The second symmetry of M-theory which should be reproduced by Matrix
theory is the group of duality symmetries of the theory.  We now
discuss these symmetries in turn.

\vspace{0.08in}
\noindent
{\it \lsm\ \ \ Lorentz symmetry in Matrix theory}
\vspace{0.05in}

There is as of yet very little evidence for a residual Lorentz
symmetry in Matrix theory.  In fact, this is one of the directions in
which the least progress has been made.  Some evidence that the theory
has Lorentz invariance at the classical level was given by de Wit,
Marquard and Nicolai \mcite{dmn}.  As stressed by BFSS, however, the
quantum version of this argument is liable to be much more subtle.
Other results relevant to the Lorentz symmetry of the theory include
calculations of scattering with longitudinal momentum transfer which
we discuss further below.

\vspace{0.08in}
\noindent
{\it \tds\ \ \ Duality in Matrix theory}
\vspace{0.05in}

In addition to Lorentz symmetry, M-theory has a set of
duality symmetries which appear when the theory has been compactified
on a $d$-dimensional torus.  This group of
``U-duality'' symmetries increases in size and complexity as each
additional dimension is compactified, as  discussed in the lectures
of Mukhi in this school.  In this section we discuss the case $d = 3$
in some detail.  Compactification on tori of other dimensions is
discussed briefly in Section  \ref{sec:compactification2}.

After compactification on $T^3$, the U-duality group of M-theory is
$SL(3,Z) \times SL(2,Z)$.  This group is generated by two types of
elementary symmetries.  The $SL(3,Z)$ part of the U-duality group
corresponds to the symmetry group of the moduli space of $T^3$
compactifications.  Since this symmetry is simply related to the
compactification space, it is a manifest symmetry of toroidally
compactified Matrix theory.  The $SL(2,Z)$ part of the U-duality group
corresponds to a form of M-theory T-duality \mcite{Sen,Aharony}.
Symmetries in this group can invert the volume of the compactification
3-torus, and are not manifest from the Matrix theory point of view.
We will now discuss T-duality in M-theory and its realization in
Matrix theory.

One simple way to understand T-duality in M-theory is through its
relationship with IIA T-duality.  If we compactify M-theory on a
3-torus in dimensions 8, 9 and 11, then we can draw the following
commuting diagram of duality symmetries
\begin{center}
\centering
\begin{picture}(200, 60)(- 100,- 25)
\put(-25,15){\makebox(0,0){M}}
\put(25,15){\makebox(0,0){M}}
\put(-25, -15){\makebox(0,0){IIA}}
\put(25,-15){\makebox(0,0){IIA}}

\put(-15,15){\vector(1,0){30}}
\put(15,15){\vector(-1,0){30}}
\put(-13,-15){\vector(1,0){26}}
\put(13, -15){\vector(-1,0){26}}

\put(-25,7){\vector( 0, -1){14}}
\put(25,7){\vector( 0, -1){14}}

\put(-40, 0){\makebox(0,0){\tiny $R_{11}$}}
\put(40, 0){\makebox(0,0){\tiny $R_{11}$}}
\put(0,25){\makebox(0,0){\tiny $T_M$}}
\put(0,-6){\makebox(0,0){\tiny $T_{89}$}}

\end{picture}
\end{center}
Start with M-theory on the upper left.  After compactification on
dimension 11 (the vertical arrow) this becomes IIA on $T^2$.  Two
T-duality transformations give us another IIA theory on the dual of
the original $T^2$.  This corresponds to a new compactification of
M-theory, so that by moving around the diagram we define an
isomorphism of M-theory.  This isomorphism is an element $T_M$  of the
$SL(2,Z)$ T-duality group of M-theory.

A remarkable feature of this duality symmetry is that it acts on
M-theory in a way which is symmetric in dimensions 8, 9 and 11.  More
precisely, after exchanging dimensions 8 and 9, the action of $T_M$ on
the original compactification $T^3$ is  to invert the volume
of the torus through $T_M: V = R_8 R_9 R_{11} \leftrightarrow 1/V$.
This can be verified directly by following the various coupling
constants and radii around the diagram above.

It is interesting to consider the effects of the symmetry $T_M$ on the
various string and membrane states in the theory.  Momentum on the
original $T^3$ can be identified with an element of the lattice dual
to that defining the compactification torus.  (i.e., for each compact
direction $a$ on the torus there is a corresponding integer momentum
$k_a$.)  Similarly, a membrane which has been wrapped around some
2-cycle on $T^3$ can be identified with a vector on the dual lattice
which is perpendicular to the membrane.
M-theory T-duality exchanges these two dual vectors, swapping membrane
wrapping number with string momentum in the compact directions.  We
can easily check this in various special cases by following the T-duality
symmetry through the above diagram.  For example, if we begin with an
M-theory membrane wrapped in directions 9 and 11, after projection
into IIA this becomes a string wrapped on dimension 9.  Two IIA
T-dualities take this into a string with momentum in dimension 9 and
no winding.  Exchanging dimensions 8 and 9 turns this into momentum in
dimension 8.  This lifts back into momentum in dimension 8 in
M-theory, which is a  vector 
orthogonal to the original 9-11 membrane.  The reader
can check as an exercise that an 8-9 membrane in M-theory is mapped
into a state with 11-momentum in a similar fashion.

Now that we have discussed M-theory T-duality in some detail, we can
ask how this symmetry is realized in Matrix theory.  We would expect
that if Matrix theory is compactified on a 3-torus, say in
dimensions 7, 8 and 9, then the theory should have 
an $SL(2,Z)$ group of self-duality symmetries corresponding to the
group of M-theory T-dualities.  From the discussion of
compactification in Section \ref{sec:compactification}, we expect that
Matrix theory on $T^3$ should correspond to ${\cal N} = 4$ super
Yang-Mills theory on the dual $\hat{T}^3$.  In fact, this theory does have a
nontrivial $SL(2,Z)$ self-duality symmetry: the S-duality symmetry discussed in
Section \ref{sec:S-duality}.  This is precisely the duality symmetry which
implements the Matrix theory version of M-theory T-duality
\mcite{Susskind-duality,grt}.

As evidence for this identification of ${\cal N} = 4$
super Yang-Mills S-duality with
Matrix theory T-duality, we can consider the following observations:
First, as discussed in Section \pbt, the Yang-Mills
coupling constant for Matrix theory on $T^3$ is given by
\begin{equation}
\tau = \frac{i}{g_{{\rm YM}}^2}  \sim iV_{789}.
\end{equation}
Under one element of the SYM S-duality group this coupling constant is
inverted through $\tau \rightarrow -1/\tau$; this corresponds to the
inversion of the volume of the torus which we expect from the element
$T_M$ of the M-theory T-duality group.  Second, SYM S-duality
exchanges electric and magnetic fluxes.  We have identified membranes
in Matrix theory with magnetic flux in the corresponding SYM theory
\begin{equation}
{\rm Tr}\;[X^a, X^b] \sim \int iB^{ab}
\end{equation}
and
momentum in Matrix theory with electric flux through
\begin{equation}
{\rm Tr}\;  \Pi^a = {\rm Tr}\;  \dot{X^a}  \sim
\int {\rm Tr}\; \dot{A_a} = \int {\rm Tr}\; E^a.
\end{equation}
The exchange of these quantities corresponds precisely to the exchange
of membrane winding and momentum expected of M-theory T-duality.

Thus, although S-duality in ${\cal N} = 4$ super Yang-Mills has not
yet been definitively proven, we have strong evidence that Matrix
theory has the expected T-duality symmetry of M-theory, and that it
can be expressed precisely in terms of this more widely accepted field
theory duality symmetry.  Combining the $SL(2,Z)$ of SYM S-duality with the
manifest $SL(3,Z)$ symmetry of the 3-torus we find that Matrix theory
has the full U-duality group expected from M-theory compactified on a
3-torus.

\subsection{Matrix theory objects}

We will now discuss evidence that Matrix theory contains most or all
of the objects which we expect to see in the 11-dimensional
supergravity theory which is the low-energy
limit of M-theory.

\vspace{0.08in}
\noindent
{\it \mog\ \ \ Supergravitons}
\vspace{0.05in}

Let us first discuss the appearance of supergravitons in Matrix
theory.  Since 0-branes are the carriers of longitudinal momentum, we
would expect a supergraviton with longitudinal momentum $N/R$ to
correspond to a bound state of $N$ 0-branes.  From the fact that
Matrix theory has 16 supersymmetries, we know that threshold bound
states of 0-branes must live in a 256-dimensional representation of
the supersymmetry algebra.  This corresponds precisely to the number
of Kaluza-Klein modes of the supergraviton arising from the graviton,
3-form, and gravitino (256 = 44 + 84 + 128).  It has been shown that
these bound states exist, at least for $N$ prime
\mcite{Sethi-Stern,Yi,Porrati-Rozenberg}.

One remarkable feature of Matrix theory which is worth emphasizing at
this point is that
second quantization  is {\it automatic} in Matrix theory.  That is,
not only does Matrix theory naturally contain a set of states
corresponding to single gravitons, it actually has a Hilbert
space containing states with arbitrary numbers of separated
gravitons.  The point is that in the large $N$ limit we can have
matrices which break up into an arbitrary number of blocks.  For
example, a state with the schematic form
\begin{equation}
\left(\begin{array}{cccc}
M_1^a & 0 & 0 & \ddots\\
0 & M_2^a& \ddots & 0\\
0 & \ddots & \ddots & 0\\
\ddots & 0 & 0 & M_k^a
\end{array} \right)
\end{equation}
could describe a state of $k$ supergravitons, where the matrices $M_i$
are $N_i \times N_i$ matrices and the longitudinal momentum of the
$i$th graviton is $p_{+}= N_i/R$.  A matrix of this form, of course,
corresponds to a classical Matrix theory configuration.  A quantum
state describing multiple separated gravitons would be described by a
wavefunction which would approximate the tensor product of a number of
bound state wavefunctions as the separations between the gravitons  are
taken to be very large.

\vspace{0.08in}
\noindent
{\it \mom\ \ \ Supermembranes}
\vspace{0.05in}

We now discuss the appearance of the supermembrane in Matrix theory.
It was realized many years ago that there is a remarkable connection between
matrix quantum
mechanics and the light-front supermembrane
\mcite{Goldstone-membrane,Hoppe-all,bst2,dhn}.  From the point of
view that has been taken in these notes, the easiest way to see that
the supermembrane must appear in Matrix theory is to note that the
(unwrapped) supermembrane corresponds to the 2-brane of type IIA
string theory.  As discussed in Section \ref{sec:branes-smaller}, when
the theory is compactified on a 2-torus of area $A$, a 2-brane can be
built from 0-branes by constructing a 0-brane configuration with
\begin{equation}
{\rm Tr}\;[X^1, X^2] = \frac{iA}{2 \pi}.
\label{eq:membrane-torus}
\end{equation}
The energy of this configuration is 
\begin{equation}
E = -\frac{1}{2 R}  {\rm Tr}\;[X^1, X^2]^2
= \frac{A^2}{8 \pi^2 R N} 
= \frac{A^2}{32 \pi^4 \alpha'^2 R N},
\end{equation}
where factors of $2 \pi \alpha'$ have been restored in the final expression.
This corresponds to the second term in an expansion in $1/N$ of the Born-Infeld
energy for a system of $N$ 0-branes and a single 2-brane
\begin{equation}
E_{{\rm BI}} = \sqrt{(N \tau_0)^2+ (A \tau_2)^2}
= N \tau_0 + \frac{A^2 \tau_2^2}{2 N \tau_0} + \cdots.
\end{equation}
Rewritten in Planck units the energy is
\begin{equation}
E= \frac{A^2 R}{32 \pi^4 N} 
\end{equation}
which is precisely the light-front energy $E = (T_2A)^2/2p^+$ for an
M-theory membrane with area $A$, tension $T_2 = 1/(2 \pi)^2$ and
longitudinal momentum $p^+$.

To describe an infinite flat supermembrane in the noncompact theory, we 
can consider a pair of infinite matrices $X^1, X^2$ satisfying
\begin{equation}
[X^1,  X^2] = \frac{i}{2 \pi\rho} \identity
\label{eq:membrane-infinite}
\end{equation}
For example, these matrices could be taken to be proportional to the
operators $q, p = -i \;d/dq$ acting on wave functions in one dimension
\mcite{BFSS}.  Comparing to (\ref{eq:membrane-torus}) we see that $\rho
\sim N/A$ corresponds to the density of 0-branes (longitudinal
momentum) on the membrane.  Note that (\ref{eq:membrane-infinite})
cannot be satisfied by any finite dimensional matrices, but has
solutions only in the large $N$ limit.

In addition to flat membranes which are either infinite or wrapped
around a compact direction, it is desirable to have a Matrix theory
description of finite-size compact membranes moving in a noncompact
space.  In fact, precisely such configurations were described in the
work of de Wit, Hoppe and Nicolai almost a decade ago \mcite{dhn}.
These authors studied the supermembrane theory in eleven dimensions in
light-front coordinates.  In light-front gauge, the supermembrane
theory has a residual invariance under the group of area-preserving
diffeomorphisms on the world-volume.  This group can be identified as
a large $N$ limit of $SU(N)$.  This leads to a discretization of the
supermembrane theory which gives precisely the 0-brane quantum
mechanics theory.  The key ingredient in the derivation of this result is the
construction of an explicit correspondence between functions on the
membrane and matrices in $U(N)$ \mcite{Goldstone-membrane,Hoppe-all}.
In the case of a membrane of spherical topology, this correspondence
is particularly simple; functions on the 2-sphere which are expressed
in terms of polynomials in the euclidean coordinates $x_1, x_2, x_3$
are described in Matrix theory by the equivalent symmetrized
polynomials in the generators $J_1, J_2, J_3$ of the $N$-dimensional
representation of $SU(2)$.  As a simple example, we can consider the
matrix representation of a symmetric 2-sphere.  A rotationally
invariant 2-sphere of radius $r$ can be embedded in the first three
transverse directions of space through
\begin{equation}
X_a = \frac{2r}{N}  J_a, \;\;\;\;\;  a \in \{1, 2, 3\}.
\end{equation}
Even at finite $N$ this matrix configuration has a number of
geometrical properties which are associated with  a smooth 2-sphere
\mcite{Dan-Wati}.  For example, the matrices $X_a$ satisfy
\begin{equation}
X_1^2 + X_2^2 + X_3^2 = r^2 \identity +{\cal O} (1/N^2)
\end{equation}
so that in a noncommutative sense the component 0-branes are
constrained to lie on the 2-sphere.  This construction of the Matrix
theory spherical membrane is closely related to 
the ``fuzzy'' 2-sphere which appears
in mathematical work on noncommutative geometry
\mcite{Madore,Madore-book}.
Toroidal Matrix theory membranes are similarly related to the fuzzy
torus \mcite{BFSS}.

\vspace{0.08in}
\noindent
{\it \molf\ \ \ Longitudinal 5-branes}
\vspace{0.05in}

We now discuss 5-branes in Matrix theory.  There are two ways in which
the M-theory 5-brane can appear as an object in Matrix theory.  On the
one hand, it can be wrapped around the longitudinal direction, in
which case it appears as a 4-brane in Matrix theory.  On the other
hand, it can be unwrapped in the longitudinal direction in which case
it should appear as a true (NS) 5-brane in Matrix theory.  We will discuss
both cases, but we begin with the longitudinal 5-brane (L5-brane).

Longitudinal 5-branes in Matrix theory were first discussed by
Berkooz and Douglas \mcite{Berkooz-Douglas}.  They included these
branes as backgrounds for the 0-brane quantum mechanics theory by including
hypermultiplets in the theory corresponding to 0-4 strings.  In this
work the L5-branes did not appear as dynamical objects described in
terms of matrix variables.  The authors showed, however, that a
membrane which is moved around the L5-brane in the background will
pick up a Berry's phase which corresponds with that expected from the
effects of the 3-form field in supergravity.

A description of L5-branes in terms of Matrix theory variables can
be given in a fashion directly analogous to the above discussion of
the membrane \mcite{grt}.  If we compactify on a $T^4$ of volume $V$
then as discussed in Section \ref{sec:branes-smaller} a flat 4-brane
wrapped around the torus can be constructed from a set of matrices
satisfying
\begin{equation}
{\rm Tr}\;\epsilon_{abcd} X^a X^b X^c X^d = \frac{V}{2 \pi^2} 
\end{equation}
Taking the large volume limit of the torus, a
construction of a noncompact 4-brane with longitudinal momentum
density $N/V = \rho$ can be given in terms of infinite matrices
satisfying
\begin{equation}
\epsilon_{abcd} X^a X^b X^c X^d  = \frac{1}{2 \pi^2 \rho}  \identity.
\end{equation}
There are a number of ways of
constructing a configuration of this type.  One can construct a
``stack of 2-branes'' solution with 2-brane charge as well as 4-brane
charge \mcite{bss}.  It is also possible to construct a
configuration with no 2-brane charge by identifying $X^a$ with the
components of the covariant derivative operator for an instanton on $S^4$
\begin{equation}
X^a = i \partial^a + A_a.
\end{equation}
This construction is known as the Banks-Casher
instanton \mcite{Banks-Casher}.

Just as for the membrane, it is possible to construct a matrix
configuration corresponding to an L5-brane which has the transverse
geometry of a symmetric 4-sphere \mcite{clt}.  A spherical
configuration corresponding to $n$ superimposed L5-brane spheres with
radius $r$ is defined through
\begin{equation}
X_a = \frac{r}{n}  G_a, \;\;\;\;\;  a \in \{1,  \ldots,  5\}.
\label{eq:sphere}
\end{equation}
where $G_a$ are the generators of the $n$-fold symmetric tensor
product representation of the five four-dimensional gamma matrices
$\Gamma_a$.  Although these configurations have the geometrical and
physical properties expected of $n$ coincident L5-brane spheres, they
also have a number of surprising characteristics.  These
configurations can only be defined for $N$ of the form
\begin{equation}
N =\frac{(n + 1) (n + 2) (n + 3)}{6}.
\end{equation}
Furthermore, unlike the case of the membrane 2-sphere where arbitrary
fluctuations can be described by symmetrized polynomials in the
generators $J_a$, it seems that no similar approach correctly describes
fluctuations around the symmetric 4-sphere configuration.  This
Matrix 4-sphere is closely related to the
fuzzy 4-sphere which has been discussed in the context of
noncommutative geometry \mcite{gkp}.

\vspace{0.08in}
\noindent
{\it \motf\ \ \ Transverse 5-branes}
\vspace{0.05in}

A systematic way of understanding the membrane and 5-brane charges in
Matrix theory arises from considering the supersymmetry algebra of the
theory.  Schematically, the 11-dimensional supersymmetry algebra takes
the form
\begin{equation}
\{Q, Q\} \sim P^I + Z^{I_1 I_2} + Z^{I_1 \ldots I_5}
\end{equation}
where the central terms correspond to 2-brane and 5-brane charges.
The supersymmetry algebra of Matrix theory was explicitly computed by
Banks, Seiberg and Shenker \mcite{bss}.  Similar calculations had been
performed previously \mcite{Claudson-Halpern,dhn}; however, in these
earlier analyses terms such as ${\rm Tr}\;[X^a, X^b]$ and ${\rm
Tr}\;X^{[a} X^{b} X^c X^{d]}$ were dropped since they vanish for
finite $N$.  The full supersymmetry algebra of the theory takes the
form
\begin{equation}
\{Q, Q\} \sim P^I + z^a + z^{ab} + z^{abcd}.
\end{equation}
The charge
\begin{equation}
z^{ab} \sim {\rm Tr}\;[X^a, X^b]
\end{equation}
corresponds to membrane charge.
The charge
\begin{equation}
z^{abcd} \sim {\rm Tr}\;X^{[a} X^{b} X^c X^{d]}
\end{equation}
corresponds to longitudinal 5-brane charge, as we have just discussed.
The charge
\begin{equation}
z^a \sim {\rm Tr}\; \{P^b,[X^a, X^b]\}
\end{equation}
corresponds to longitudinal membranes (strings).  This can be
understood easily in a dual Yang-Mills picture, where this charge
corresponds to the Poynting vector $F^{ab}E^b$; as usual, momentum is
the dual of string winding number.

Nowhere in this analysis of brane charges do we see any sign of a
charge corresponding to transverse 5-branes.  As we will  see in
the next section, there is also no sign of such a charge in the
general expression for the leading long-range gravitational
interaction between two matrix objects \mcite{Dan-Wati2}.  It was
argued by Banks, Seiberg and Shenker that in fact transverse 5-branes
cannot exist in the IMF since they are Dirichlet objects for the
M-theory membrane \mcite{bss}.  Nonetheless, there is an argument
\mcite{grt} that a T5-brane can be constructed implicitly using the
super Yang-Mills S-duality of Matrix theory on $T^3$.  We now review
this argument briefly.

Let us compactify M-theory on dimensions 7, 8 and 9.  We  now place
an infinite membrane along dimensions 5 and 6.  Performing M-theory
T-duality on dimensions 7, 8 and 9 has the effect of taking the
membrane to a 5-brane wrapped around dimensions 5-9, as can be seen in
the following commuting diagram
\begin{center}
\centering
\begin{picture}(200, 60)(- 100,- 25)
\put(-25,15){\makebox(0,0){M}}
\put(25,15){\makebox(0,0){M}}
\put(-25, -15){\makebox(0,0){IIA}}
\put(25,-15){\makebox(0,0){IIA}}
\put(-50, 15,){\makebox(0,0){\small M2(56)}}
\put(-50, -18,){\makebox(0,0){\small D2(56)}}
\put(55,15){\makebox(0,0){\small M5(56789)}}
\put(55,-18){\makebox(0,0){\small D4(5678)}}

\put(-15,15){\vector(1,0){30}}
\put(15,15){\vector(-1,0){30}}
\put(-13,-15){\vector(1,0){26}}
\put(13, -15){\vector(-1,0){26}}

\put(-25,7){\vector( 0, -1){14}}
\put(25,7){\vector( 0, -1){14}}

\put(-40, 0){\makebox(0,0){\tiny $R_{9}$}}
\put(40, 0){\makebox(0,0){\tiny $R_{9}$}}
\put(0,25){\makebox(0,0){\tiny $T_M$}}
\put(0,-6){\makebox(0,0){\tiny $T_{78}$}}

\end{picture}
\end{center}

Thus, to construct a T5-brane in Matrix theory we must begin with the
theory compactified on $T^3$ in dimensions 7-9, with a Yang-Mills
configuration having scalar fields satisfying
\begin{equation}
[X^5, X^6] \sim i\identity
\label{eq:T5-brane}
\end{equation}
Performing SYM S-duality on this state should give a transverse
5-brane.  There are a number of puzzling subtleties regarding
this construction, however.  First, we have no explicit representation of
S-duality in 4D SYM, so we cannot construct the T5-brane state
explicitly.  Second, there is a confusing issue about how the large
$N$ limit must be taken.  In order to construct a configuration like
(\ref{eq:T5-brane}) we must take the large $N$ limit before performing
the S-duality transformation.  It is unclear how SYM S-duality behaves
in the large $N$ limit.  Finally, if this state truly exists, a good
reason needs to be found why the corresponding charge does not appear
in the supersymmetry algebra or in the leading term in the
long-distance potential.  It is possible that this charge may be
nonlocal, and vanishes for a reason analogous to the vanishing of the
L5-brane and membrane charges at finite $N$.

\subsection{Interactions in Matrix theory}
\label{sec:matrix-interactions}

\vspace{0.08in}
\noindent
{\it \rss\ \ \ The leading $1/r^7$ potential}
\vspace{0.05in}

We now turn our attention to the interactions between the objects of
Matrix theory.  We discussed in Section \sczz\ the calculation of the
velocity-dependent effective potential (\ref{eq:scattering-potential})
between a pair of 0-branes in super Yang-Mills quantum mechanics.
This potential was found as the result of a one-loop calculation in
the Yang-Mills theory.  As pointed out by BFSS \mcite{BFSS}, this
potential corresponds precisely with the leading long-range
supergravity potential between a pair of gravitons with longitudinal
momentum $1/R$ in light-front coordinates.  An explicit calculation
shows that the leading term in the long-range supergravity potential
between a pair of pointlike objects with momenta $\hat{p}^{I}$ and
$\tilde{p}^{I}$ due to the exchange of a single graviton with no
longitudinal momentum is (in string units)
\begin{equation}
V_{\rm gravity} = - {15 \over 4} \, {R^4 \over r^7}
\left[\left(\hat{p} \cdot \tilde{p}\right)^2 - {1 \over 9}  \hat{p}^2
\tilde{p}^2\right].
\label{eq:graviton-potential}
\end{equation}
Taking one of the Matrix theory 0-branes to be at rest and the other to have
transverse velocity $v^a$, we have
\begin{eqnarray}
\hat{p}^+ = \frac{1}{R}  \;\;\;\;\; & &  \hat{p}^a = 0 \;\;\;\;\;
\;\;\;\;\; \;\;\;
\hat{p}^-= 0\\
\tilde{p}^+ = \frac{1}{R}  \;\;\;\;\; & &  \tilde{p}^a = \frac{v^a}{R} 
 \;\;\;\;\; \;\;\;\;\;
\tilde{p}^-=  \tilde{p}^2_\perp/2 \tilde{p}^+ =  \frac{v^2}{2 R} \nonumber
\end{eqnarray}
Inserting these momenta into (\ref{eq:graviton-potential}) gives
\begin{equation}
V_{{\rm gravity}} = -\frac{15v^4}{16 \;r^7} 
\end{equation}
in exact agreement with (\ref{eq:scattering-potential}).  This exact
correspondence carries through for states with longitudinal momentum
$p^+ = N/R$.  The gravitational potential in this case is simply
multiplied by the product $\hat{N} \tilde{N}$.  The same factor enters
the Yang-Mills calculation because this is the multiplicity of string
states stretching between the two collections of 0-branes, assuming
that each of the two states is described by a localized bound state of
$N$ 0-branes.

In the last year or so, the one-loop potential has been calculated for
a variety of Matrix theory objects
\mcite{DKPS,BFSS,Aharony-Berkooz,Lifschytz-Mathur,Lifschytz-46,bc,Lifschytz-transverse,ChepTseyI,MaldacenaI,Vakkuri-Kraus,Gopakumar-Ramgoolam,Chepelev-Tseytlin2,MaldacenaII,Esko-Per2}.
In all cases this potential was found to agree at leading order with
the expected leading long-distance potential from supergravity.  A
general proof of this result was given by Kabat and the author in
\mycite{Dan-Wati2}; we will now describe briefly the analysis in the
general case.  Given a pair of classical Matrix theory objects
described by matrices $\hat{X}$ and $\tilde{X}$ of sizes
$\hat{N}\times\hat{N}$ and $\tilde{N}\times\tilde{N}$ respectively,
the one-loop potential between these objects can be calculated in the
quasi-static appropriation by taking a background configuration
\begin{equation}
\langle X^a \rangle =
\left(\begin{array}{cc}
\hat{X}^a & 0\\
0 & \tilde{X}^a
\end{array} \right).
\end{equation}
Summing the frequencies of the string oscillators associated with the
bosons, fermions and ghosts in the off-diagonal matrix blocks as in
(\ref{eq:oscillator-sum}) gives the one-loop potential between the two
objects.  If the centers of mass of the two objects are separated by a
distance $r$ which is large compared to the sizes of the objects then
this potential can be expanded in powers of $1/r$.  The leading term
is of the form $F^4/r^7$ where $F$ can be a term of the form $\dot{X}$
or $[X, X]$ \mcite{Metsaev-Tseytlin,Dan-Wati}.  Decomposing the general
expression for this term into functions of $\hat{X}$ and $\tilde{X}$,
and grouping terms by their Lorentz structure it can be shown
\mcite{Dan-Wati2} that the leading term in the Matrix theory potential
between an arbitrary pair of separated objects is given by
\begin{equation}
\label{eq:matrix-potential-general}
V_{\rm matrix}=V_{\rm gravity} + V_{\rm electric} + V_{\rm magnetic} 
\end{equation}
where
\begin{eqnarray}
V_{\rm gravity} & = & - {15 R^2 \over 4 r^7} \left( {\hat{\cal
T}}^{IJ} \tilde{{\cal T}}_{IJ} 
- {1 \over 9} {\hat{\cal T}}^I{}_I \tilde{{\cal T}}^J{}_J\right) 
\nonumber \\
V_{\rm electric} & = & - {45 R^2 \over r^7} {\hat{\cal J}}^{IJK}
{\tilde{\cal J}}_{IJK}  
\label{eq:matrix-interactions} \\
V_{\rm magnetic} & = & - {45 R^2\over r^7} \hat{{\cal M}}^{+-ijkl}
\tilde{{\cal M}}^{-+ijkl} 
\nonumber
\end{eqnarray}
and where we define the following quantities:
$\itt^{IJ}$ is a
symmetric tensor with components
\begin{eqnarray}
\itt^{--} & = & {1 \over R} \; \str  \left( \frac{1}{4} \dot{X}^a \dot{X}^a
\dot{X}^b \dot{X}^b+
{1 \over 4} \dot{X}^a \dot{X}^a F^{bc} F^{bc}  +
 \dot{X}^a \dot{X}^b F^{ac} F^{cb} \right. \nonumber\\ 
& & \qquad \qquad  \left.  +
{1 \over 4} F^{ab} F^{bc} F^{cd} F^{da}  -
\frac{1}{16} F^{ab} F^{ab}  F^{cd} F^{cd}  \right) \nonumber
  \\
\itt^{-a} & = & {1 \over R} \;\str \left(\frac{1}{2} \dot{X}^a \dot{X}^b \dot{X}^b +
{1 \over 4} \dot{X}^a F^{bc} F^{bc} 
                  + F^{ab} F^{bc} \dot{X}^c \right) \nonumber \\
\itt^{+-} & = & {1 \over R} \;\str \left(\frac{1}{2} \dot{X}^a \dot{X}^a + {1
\over 4} F^{ab} F^{ab} \right) \label{eq:matrix-t} \\ 
\itt^{ab} & = & {1 \over R}  \;\str \left( \dot{X}^a \dot{X}^b + F^{ac}
F^{cb} \right) \nonumber \\ 
\itt^{+a} & = & {1 \over R} \;\str \dot{X}^a \nonumber \\
\itt^{++} & = & {1 \over R} \;\str \identity =
{N \over R} \nonumber
\end{eqnarray}
$\ijj^{IJK}$ is a totally antisymmetric tensor with components
\begin{eqnarray}
\ijj^{-ab} & = & {1 \over 6 R} \str \left( \dot{X}^a \dot{X}^c F^{cb} -
\dot{X}^b \dot{X}^c F^{ca} - \frac{1}{2} \dot{X}^c \dot{X}^c F^{ab}
\right.\nonumber\\ 
& & \qquad \qquad  \left. + {1 \over 4} F^{ab} F^{cd} F^{cd} +F^{ac}
F^{cd} F^{db} \right) \nonumber \\ 
\ijj^{+-a} & = & {1 \over 6 R} \str \left( F^{ab} \dot{X}^b \right) 
\label{eq:matrix-j} \\
\ijj^{abc} & = & - {1 \over 6 R} \str \left( \dot{X}^a F^{bc} +
\dot{X}^b F^{ca} + \dot{X}^c F^{ab} \right) \nonumber \\
\ijj^{+ab} & = & - {1 \over 6 R} \str F^{ab} \nonumber
\end{eqnarray}
and
$\imm^{IJKLMN}$ is a totally antisymmetric tensor  with
\begin{equation}
\label{eq:matrix-m}
 \imm^{+-abcd} = {1 \over 12 R} \str \left(F^{ab} F^{cd} + F^{ac}
F^{db} + F^{ad} F^{bc}\right)\,.
\end{equation}
We have defined $F_{ab} = - i[X_a, X_b]$.  The trace $\str$ is defined
to be the trace symmetrized over all possible orderings of the factors
$\dot{X}$ and $F$.  Tensors $\hat{{\cal T}}^{IJ}, \tilde{{\cal
T}}^{IJ}$, etc. are defined through
(\ref{eq:matrix-t}-\ref{eq:matrix-m}) as functions of $\hat{X}$ and
$\tilde{X}$.  Note that the only components of $\imm$ which appear in
(\ref{eq:matrix-interactions}) are those defined in
(\ref{eq:matrix-m}); there is no expression known for other components
of this tensor.

The general form of the Matrix theory potential
(\ref{eq:matrix-potential-general}) can be compared with the leading
long-distance potential in 11D supergravity between an arbitrary pair
of objects.  In supergravity this potential arises from the exchange
of a single particle, which must be either a graviton or 3-form
quantum.  In light-front coordinates the propagator for a quantum
carrying no longitudinal momentum carries a factor of $\delta
(\hat{x}^+ - \tilde{x}^+)$ \mcite{Hellerman-Polchinski}.  Thus, we
expect the light-front 
supergravity potential to be instantaneous in light-front time.  An
explicit calculation of this potential shows that the leading term is
precisely given by (\ref{eq:matrix-potential-general}) where $\itt$,
$\ijj$ and $\imm$ are the integrated stress tensor, membrane current
and 5-brane current in supergravity.  Thus, if we use
(\ref{eq:matrix-t}), (\ref{eq:matrix-j}) and (\ref{eq:matrix-m}) as
definitions of the integrated stress tensor and currents in Matrix
theory, we see that there is an exact correspondence between the
leading term in the one-loop Matrix theory potential and the leading
term in the supergravity potential for an arbitrary pair of objects.

\vspace{0.08in}
\noindent
{\it \rgs\ \ \ Further aspects of Matrix theory interactions}
\vspace{0.05in}

Although the correspondence between Matrix theory and supergravity
interactions has been demonstrated in general at order $1/r^7$, the
current understanding of subleading terms is much less developed.
There are a number of ways in which subleading terms appear in the
Matrix theory potential.  A systematic analysis of the structure of
the subleading terms in the graviton scattering amplitude was carried
out by Becker, Becker, Polchinski and Tseytlin (BBPT) \mcite{bbpt} and has
also been considered by Susskind.  Generally, at $n$th loop order,
there are terms in the potential of order $v^{4 + 2k}/r^{4 + 3n + 4k}$
for all values of $k$.  For more general Matrix theory objects, the
structure is similar, with $F$ playing the role of $v$; however, there
are also dependencies on the fields $X$ so that the full expansion is
of the form
\begin{equation}
V = \sum_{n, k, l}  V_{nkl} \frac{F^{4 + 2k} X^l}{r^{4 + l + 3n + 4k}} 
\label{eq:general-expansion}
\end{equation}
where $n$ indicates the loop order at which a given term arises.
Generally, the contraction of the indices of $F$ and $X$ can be
carried out in many inequivalent ways; the coefficient $V_{nkl}$
therefore is shorthand for many independent coefficients at each order,
corresponding to all possible contractions.  We will now discuss some
of the  features of this loop expansion which are currently
understood.  Note that  our discussion focuses on interactions between
purely bosonic states.  When fermions are included there can be
additional effects such as spin effects; it has been found that these
effects seem to be captured accurately by  Matrix theory also, at
least at leading order \mcite{Harvey-spin,mss,Kraus-spin}.

If we consider only the  terms in the one-loop expansion which contain four
powers of $F$, the expansion reduces to a sum of terms of the form
$F^4 X^l/r^{7 + l}$.  This set of terms was analyzed in
\mycite{Mark-Wati,Dan-Wati2}, where it was shown that these terms
can be described in terms of interactions between higher moments of the
Matrix theory stress tensor (\ref{eq:matrix-t}), membrane current
(\ref{eq:matrix-j}) and L5-brane current  (\ref{eq:matrix-m}).  These
interactions correspond precisely to the higher-order terms expected
from supergravity for the interaction between two extended objects due
to single graviton or 3-form exchange.  It seems reasonable to
conclude that the role of the factors $X^l/r^l$ will in general be to
incorporate higher moments of
extended objects; thus, to understand the remaining terms in
(\ref{eq:general-expansion}) it will suffice to restrict attention to
the terms with $l = 0$, which are the only ones contributing in the
case of graviton scattering.

One set of terms of particular interest are the terms of the form $F^4/r^{4 +
3n}$.  If such terms existed with nonvanishing coefficients beyond
one-loop order they would renormalize the $v^4$ interaction term which
already agrees at one-loop order with supergravity.  It was
conjectured by BFSS that no such renormalization occurs.  Becker and Becker
have performed the calculation explicitly for graviton scattering at
two-loop order and shown that the term of order $v^4/r^{10}$ vanishes
identically \mcite{Becker-Becker}.  This supports the hypothesis that
all remaining $v^4$ terms vanish; however, this has not been shown at
higher order.  It has also been suggested that the $v^4$ terms may in
fact be
renormalized at higher loop order since analogous renormalizations
occur in three-dimensional theories \mcite{Dine-Seiberg}.  These terms
may also be affected by processes with longitudinal momentum transfer,
which we discuss briefly below.

The next several terms which contribute at two loops have also been
calculated for graviton scattering.  It was shown by BBPT \mcite{bbpt}
that the term of order $v^6/r^{14}$ is also in agreement with the
potential expected from classical supergravity.  This term corresponds
to a general relativistic correction to the lowest order term in the
potential.  All the terms in the Matrix theory potential of the form
$v^{4 + 2m}/r^{7m}$ carry integral powers of the gravitational
constant when expressed in Planck units.  It is believed that these
terms should reproduce classical supergravity to all orders.  Although
none of these terms have been calculated precisely beyond two loops,
it has been argued
\mcite{Chepelev-Tseytlin2,Esko-Per2,Balasubramanian-gl,Chepelev-Tseytlin3}
that the general form of these terms should correspond  with higher
order terms in the non-abelian Born-Infeld action.  Although these
terms cannot be determined uniquely by this ansatz, in certain cases
exact expressions for the supergravity interactions are of the
Born-Infeld form, and are in agreement with this conjecture.

At each loop order there are terms which contribute at higher order in
$1/r$ than the terms which are expected to correspond with classical
supergravity.  It is believed that these terms represent quantum
gravity corrections.  There has been some discussion of this question
in the literature
\mcite{Susskind-talk,Berglund-Minic,Serone,Esko-Per-short,Balasubramanian-gl,Beckers-graviton},
but as yet there does not seem to be a detailed understanding of this
issue.

The loop expansion in Matrix theory which we have been discussing only
describes processes in which no longitudinal momentum is exchanged.
Clearly, for a full understanding of interactions in Matrix theory it
will be necessary to include processes with longitudinal momentum
transfer.  Some progress has been made in this direction.  Polchinski
and Pouliot have calculated the scattering amplitude for two 2-branes
for processes in which a 0-brane is transferred from one 2-brane to
the other \mcite{Polchinski-Pouliot}.  In the Yang-Mills picture the
incoming and outgoing configurations in this calculation are described
in terms of a $U(2)$ gauge theory with a scalar field taking a VEV
which separates the branes, as discussed in Section \szt.  The transfer
of a 0-brane corresponds to an instanton-like process where a unit of
flux is transferred from one brane to the other.  The results of this
calculation are in agreement with expectations from supergravity.
This result suggests that processes involving longitudinal momentum
transfer may be correctly described in Matrix theory.  Note, however,
that the Polchinski-Pouliot calculation is not precisely a calculation
of membrane scattering with longitudinal momentum transfer in Matrix
theory since it is carried out in the 2-brane gauge theory language.
In the T-dual Matrix theory picture the process in question
corresponds to a scattering of 0-branes in a toroidally compactified
space-time with the transfer of membrane charge.  This process was
studied further by Dorey, Khoze and Mattis \mcite{dkm} and was related
to graviton scattering by Banks, Fischler, Seiberg and Susskind \mcite{BFSS2}.

\section{Matrix theory: Further Developments}
\label{sec:developments}

In the last year there has been a veritable explosion of Matrix theory
related papers.  In this section we describe briefly a few of the
interesting directions in which this work has gone.  There are a
number of important and interesting developments which we do not
discuss here at all.  In particular, nothing is said in these notes
about the recent developments on Matrix theory black holes or light-front
5-brane theories.  We also do not discuss Matrix theory
compactification on orbifolds.
Although we do give a brief description of the DVV
formulation of Matrix string theory, there are many other interesting
formulations of string theory in the matrix language, such as
heterotic Matrix strings, which we do not cover here.  Another particularly
interesting set of developments involves the compactification of
Matrix theory on higher dimensional tori.  Aside from a few brief
comments, this topic is not covered.  Some of these topics are covered
in more detail in the reviews \mycite{banks-review,Susskind-review}.

\subsection{Matrix string theory}

An interesting feature of Matrix theory is that with a few minor
modifications it can be used to give a nonperturbative definition of
string theory.  A number of approaches have been taken to Matrix
string theory \mcite{IKKT,Motl-string,Banks-Seiberg,DVV-string}.  In
this section we review briefly a few aspects of the Matrix string
theory approach due to Dijkgraaf, Verlinde and Verlinde (DVV)
\mcite{DVV-string,DVV-string2}.

If we consider Matrix theory compactified in dimension 9 on a circle
$S^1$, we have a super Yang-Mills theory in $(1+1)$-D on the dual
circle $\hat{S}^1$.  In the BFSS formulation of Matrix theory, this
corresponds to M-theory compactified on a 2-torus.  If we now think of
dimension 9 rather than dimension 11 as the dimension which has been
compactified to get a IIA theory, then we see that this super
Yang-Mills theory should provide a light-front description of type IIA
string theory.  Because we are now interpreting dimension 9 as the
dimension of M-theory compactification, the fundamental objects which
carry momentum $p^{+}$ are no longer 0-branes, but rather strings
with longitudinal momentum.  Thus, it is natural to interpret $N/R$ in
this super Yang-Mills theory as the longitudinal string momentum.  It
was argued by Dijkgraaf, Verlinde and Verlinde that in fact this gives
a natural corollary to the Matrix theory conjecture, namely that 2D
super Yang-Mills in the large $N$ limit should correspond to
light-front IIA string theory.

To examine this form of the conjecture in more detail, let us begin by
considering the Matrix theory Hamiltonian  (working in Planck units
and dropping factors of order unity as in \mcite{DVV-string})
\begin{equation}
H = R_{11} {\rm Tr}\; \left[
  \Pi_a \Pi_a -  [X^a, X^b]^2  +
\theta^T
\gamma_a[X^a, \theta] \right]\ .
\end{equation}
After compactification on $R_9$ we identify $X^9 \rightarrow R^9
D_\sigma$, $\Pi_9 \rightarrow R_9 \dot{A}_9\sim E_9/R_9$, where
$\sigma \in[0, 2 \pi]$ is the coordinate on the dual circle.  With
these identifications, and using $g \sim R_9^{3/2}$, the Hamiltonian
was rewritten by DVV in the form
\begin{eqnarray}
H  & = &  \frac{R_{11}}{2 \pi}  \int d \sigma \; {\rm Tr}\; \left[
 \Pi_a \Pi_a + (D_\sigma X^a)^2 + 
\theta^TD_\sigma \theta
\right.\nonumber\\
& &\hspace{1in} \left.
+
\frac{1}{g^2}  \left( E^2  - [X^a, X^b]^2 \right)
+ \frac{1}{g} \theta^T
\gamma_a[X^a, \theta]\right]\ .
\end{eqnarray}
This is essentially the form of the Green-Schwarz light-front string
Hamiltonian, with the modification that the fields are now $N \times
N$ matrices which do not necessarily commute.  This means that  the
theory automatically contains multi-string objects living in a second
quantized Hilbert space.  Furthermore, it is possible to construct
extended string theory objects in terms of the noncommuting matrix
variables, by a simple translation from the original Matrix theory
language.  We reproduce  in Table~\ref{tab:string}
a table of the extended objects in this Matrix string
theory.  The objects are listed in terms of their interpretations in
M-theory, as well as their interpretations in Matrix string
theory and associated charges in Matrix string theory.  Charges are
given only up to an overall constant.
\begin{table}[t]
\caption{Objects and their charges in Matrix string
theory\label{tab:string}}\vspace{0.4cm}
\begin{center}
\begin{tabular}{|l|l|l|}
\hline
M-theory & Matrix string object & Matrix string charge\\
\hline
$p_{11}$&$ p^+ $&$ N$\\
$p_9 $& D$0 $&$ E$\\
$p_a $&$ p_a $&$ \Pi_a$\\
M${}_{a9}\; \;$ ($a-9$ membrane) &$ w_a  \; \;$(wound string)&$
D_\sigma X^a$\\ 
M${}_{11\; 9} $&$ w_+ $&$ \Pi_a (D_\sigma X^a) \; \; (E \times B)$\\
M${}_{ab} $&D$2_{ab} $&$[X^a, X^b]$\\
M${}_{a \; 11} $&D$2_{a +} $&$ E D_\sigma X^a
+ \Pi_b[X^a, X^b]$\\
$5_{abc+9} $&D$4_{abc+}$&$ D^{[9} X^a X^b X^{c]}$\\
$5_{abcd+} $&NS$5_{abcd+} $&$ X^{[a} X^b X^c X^{d]} $\\
\hline
\end{tabular}
\end{center}
\end{table}
To verify each of the entries in this table it suffices to consider a
Matrix theory object with its known charge, and to rewrite that object
and charge in terms of the Yang-Mills description after the ``9-11
flip'' corresponding to interpreting dimension 9 as the M-theory
compactification direction.  For example, consider an M-theory
membrane wrapped in dimensions 8-9.  In BFSS Matrix theory this is a
membrane with charge $[X^8, X^9]$.  In Matrix string theory we take
$X^9 \rightarrow D_9$ so that the charge becomes $D_9 X^8$.  Since
dimension 9 is the M-theory compactification direction, this
corresponds to a string wrapped around dimension 8.  Note that the
only objects missing in this table are those corresponding to the
T5-brane in BFSS Matrix theory.  These correspond to a 4-brane or
5-brane wrapped in transverse directions in Matrix string theory.

One particularly nice feature of the DVV approach to Matrix string
theory is the way in which the individual string bits carrying a
single unit of longitudinal momentum combine to form long strings.  As
the string coupling becomes small $g \rightarrow 0$, the coefficient
of the term $[X^a, X^b]^2$ in the Hamiltonian becomes very large.  This
forces the matrices to become simultaneously diagonalizable.  Because
the string configuration is defined over $S^1$, however, the matrix
configuration need not be periodic in $\sigma$.  The matrices $X^a
(0)$ and $X^a (2 \pi)$ can be related by an arbitrary permutation.
The lengths of the cycles of this permutation determine the numbers of
string bits which combine into long strings whose longitudinal
momentum $N/R_{11}$ can become large in the large $N$ limit.  As the
coupling becomes very small, the theory therefore essentially becomes
a sigma model on $(\br^8)^N/S^N$.  The twisted sectors of this theory
correspond precisely to the sectors where the string bits are combined
in different permutations.  In this picture, string interactions
appear as vertex operators in the conformal field theory
arising as the infrared limit of the sigma model theory.  
It is not apparent, however, how such interactions are related to the
Yang-Mills description of the theory.
It would be nice to have a more direct understanding of this relationship.

Some discussion is given in the original DVV papers of the structure
of D-branes in Matrix string theory.  This is another direction
in which it would be interesting to develop the theory in further
detail.  For example, DVV suggest that a 0-brane corresponds to a
single string bit which does not become part of an extended string in
the large $N$ limit.  It would be nice if there were a natural
way in which the known properties of 0-branes could be derived from
this point of view.  Clearly, there is more to be said about the
relationship between Matrix string theory and other formulations of
light-front string theory.

\subsection{Compactification of more than three dimensions}
\label{sec:compactification2}

As discussed in Section \ref{sec:compactification}, Matrix theory
compactified on a torus of dimension $d \leq 3$ is described in terms
of super Yang-Mills theory on the dual torus.  
Compactification on $T^3$ was described in section \tds.
Compactification of the theory on $T^2$ was discussed by Sethi and
Susskind \mcite{Sethi-Susskind}.  They pointed out that as the $T^2$
shrinks, a new dimension appears whose quantized momentum modes
correspond to magnetic flux on the $T^2$.  In the limit where the area
of the torus goes to 0, an $O (8)$ symmetry appears.  This corresponds
with the fact that IIB string theory appears as a limit of M-theory on
a small  2-torus \mcite{Aspinwall-duality,Schwarz-multiplet}.
When more than three
dimensions are toroidally compactified, the theory undergoes
even more remarkable transformations \mcite{fhrs}.  For example, consider
compactifying the theory on $T^4$.  The manifest symmetry group of
this theory is $SL(4,Z)$.  The expected U-duality group of
M-theory compactified on $T^4$ is $SL(5,Z)$, however.  It was pointed out by
Rozali \mcite{Rozali} that the U-duality group can be completed by
interpreting instantons on $T^4$ as momentum states in a 
fifth compact dimension.  This means that Matrix
theory on $T^4$ is most naturally described in terms of a (5 +
1)-dimensional theory with a chiral $(2, 0)$ supersymmetry.  This $(2,
0)$ theory is an unusual theory with 16 supersymmetries
\mcite{Seiberg-16} which appears to play a crucial role in a wide
variety of properties of M-theory and 5-branes.

Compactification on tori of higher dimensions than four continues to
lead to more complicated situations, particularly in the case of
$T^6$.  A significant amount of literature has been produced on this
subject, to which the reader is referred to further details (see for
example \mycite{Sen,Seiberg-DLCQ,Berkooz-duality,Ganor-Sethi} and
references therein).  Despite the complexity of $T^6$
compactification, however,
it was recently suggested by Kachru, Lawrence
and Silverstein \mcite{kls} that compactification of Matrix theory on a more
general Calabi-Yau 3-fold might actually lead to a simpler theory than
that resulting from compactification on $T^6$.

\subsection{Proofs and counterexamples}
\label{sec:recent}

Since the time when these lectures were given, there has been a great
deal of debate about whether the Matrix theory conjecture is truly
correct to all orders in perturbation theory, and if so, why.  The
results of this debate are still uncertain, and a full discussion of
the issues involved will not be given here.  We will only briefly
review a few of the points in the discussion.

It was suggested by Susskind \mcite{Susskind-DLCQ} that there might be a sense
in which the Matrix theory conjecture holds even at finite $N$.  This
extended version of the conjecture would relate finite $N$ Matrix
theory with the finite $N$  discrete light-cone quantization of
M-theory.   An argument has been given by Seiberg \mcite{Seiberg-DLCQ}
which seems to indicate that this correspondence is correct, and
related discussions have been given by Sen \mcite{Sen} and de Alwis
\mcite{dealwis-DLCQ}.  Seiberg's approach even seems to apply to
compactification of Matrix theory on an arbitrary manifold, although
the details of this argument have not been made precise.

However,
there are also a number of pieces of evidence that the correspondence
between Matrix theory and supergravity breaks down in certain
contexts.  Attempts to formulate Matrix theory on curved
spaces seem to lead to discrepancies between the leading terms in the
Matrix theory and supergravity interaction potentials
\mcite{dos,Douglas-Ooguri}.  Higher-loop effects on orbifolds also seem
to give rise to discrepancies \mcite{ggr,ddm} (further comments on this
issue appear in \mycite{Berglund-Minic}).  Even in flat space, it seems
that Matrix theory may have problems in reproducing supergravity:
Recently Dine and Rajaraman considered 3-graviton scattering in Matrix
theory \mcite{Dine-Rajaraman}, and argued that there are certain
diagrams in supergravity with nonzero amplitudes which simply cannot
be reproduced at any order in Matrix theory.  When considering
interactions between extended objects, it also seems that Matrix
theory may diverge from supergravity in unusual ways; although the
one-loop interaction between any two objects must agree with
supergravity if the components of the source tensors are defined as in
Section \rgs, the components of the stress tensor for extended objects
are defined so that the equivalence principle seems to break down at
finite $N$ \mcite{Dan-Wati2}.

Thus, we seem to be faced with a contradiction: on the one hand proofs
that even at finite $N$ Matrix theory is a correct description of DLCQ
M-theory, on the other hand evidence that Matrix theory does not agree
with results one would expect from supergravity.  Some recent papers
have addressed this puzzle
\mcite{banks-review,Hellerman-Polchinski,Balasubramanian-gl,Bilal};
however, a complete resolution of the situation will certainly take
some time.  It may be that to resolve these issues it will be
necessary to understand the large $N$ limit of Matrix theory in a more
precise fashion.  It may also be that detailed aspects of the bound
state wave functions of gravitons will play a role in resolving these
contradictions.

\section{Conclusions}
{}

We have seen that a remarkable number of interesting properties of
D-branes can be understood from the point of view of the low-energy
super Yang-Mills description.  Super Yang-Mills theory contains
information about all the known duality symmetries of type II string
theory and M-theory.  We can construct higher- and lower-dimensional
branes of various types directly within the super Yang-Mills theory
for branes of a particular dimension.  Super Yang-Mills theory even
seems to know about supergravity.  From super Yang-Mills theory we
have some suggestion of how to generalize our notions of geometry to
include noncommutative spaces such as those of Connes
\mcite{Connes}. All this is certainly an indication that super
Yang-Mills theory is a much richer theory than it has usually been
given credit for.  Whether it will truly reproduce all of the physics
of string theory, let alone the standard model, however, remains to be
seen.  To this author, it seems that there is still some fundamental
principle lacking.  In particular, for Matrix theory to leave the IMF
or light-front gauge, it is necessary to introduce anti-0-branes.
These are virtually the only objects which cannot be constructed from
0-branes as some kind of generalized fluxes.  As has been suggested by
a number of authors, it seems that we need some more fundamental
structure in which all the objects of the theory, even 0-branes and
anti-0-branes, appear as some generalized type of fluxes or as composites
of some underlying medium.  Because of the way in which quantum
mechanics and geometrical constraints often seem to be related through
dualities, it easy to imagine that whatever fundamental principles we
are currently lacking will necessitate a rather substantial reworking
of our concepts of quantum mechanics and field theory themselves.

\section*{Acknowledgments}
I would like to thank the International Center for Theoretical Physics
in Trieste for their hospitality and for running the summer school at
which these lectures were presented.  I would also particularly like
to thank Lorenzo Cornalba, Dan Kabat, Sanjaye Ramgoolam, L\'arus
Thorlacius and Mark Van Raamsdonk for reading a preliminary draft of
these notes and making numerous helpful suggestions and corrections.
This work was supported by the National Science Foundation (NSF) under
contract PHY96-00258.

\section*{References}
\bibliographystyle{unsrt}


\end{document}